\documentclass[sigconf]{acmart}            

\newif\ifnotes
\notesfalse 

\newif\iftrack
\tracktrue

\AtBeginDocument{%
  }

\copyrightyear{2026}
\acmYear{2026}
\setcopyright{cc}
\setcctype{by}
\acmConference[CHI '26]{Proceedings of the 2026 CHI Conference on Human Factors in Computing Systems}{April 13--17, 2026}{Barcelona, Spain}
\acmBooktitle{Proceedings of the 2026 CHI Conference on Human Factors in Computing Systems (CHI '26), April 13--17, 2026, Barcelona, Spain}
\acmPrice{}
\acmDOI{10.1145/3772318.3791117} 
\acmISBN{979-8-4007-2278-3/2026/04}

\usepackage{acmart-taps}

\usepackage{xspace}
\usepackage{xcolor}
\usepackage{colortbl}
\usepackage{fontawesome5}

\newcommand{\bstart}[1]{\vspace{1mm} \noindent{\textbf{#1:}}}
\newcommand{\bpstart}[1]{\vspace{1mm} \noindent{\textbf{#1.}}}

\newcommand{\bstartnc}[1]{\noindent{\textbf{#1}}}

\definecolor{activegold}{RGB}{255,193,61}

\definecolor{lightorange}{rgb}{1,0.8,0.4}
\definecolor{lightorange}{RGB}{230, 170, 50}
\definecolor{lightgreen}{RGB}{121,210,121}
\definecolor{beige}{RGB}{204,189,128}
\definecolor{lightteal}{RGB}{121,199,210}
\definecolor{lightblue}{RGB}{100,212,239}
\definecolor{lightpurple}{RGB}{153,102,255}
\definecolor{lightred}{RGB}{245, 132, 120}
\definecolor{red}{RGB}{178,34,34}
\definecolor{gray}{RGB}{166,166,166}
\definecolor{indexBlue}{cmyk}{0.9,0.8,0,0}
\definecolor{indexGreen}{cmyk}{0.8,0.2,0.8,0.55}
\definecolor{deepblue}{cmyk}{0.9,0.75,0,0.5}
\definecolor{deepred}{cmyk}{0,0.75,0.75,0.4}
\definecolor{pink}{RGB}{214,114,0}

\usepackage{enumitem}

\definecolor{forestgreen}{RGB}{50,120,50}

\newcommand{\ie}{i.e.,\@\xspace}
\newcommand{\eg}{e.g.,\@\xspace}
\newcommand{\etal}{et~al.\@\xspace}

\newcommand{\numChallenges}{16}
\newcommand{\categoryTTT}{\textbf{Technological}}
\newcommand{\categoryPeople}{\textbf{Social}}
\newcommand{\categoryAI}{\textbf{AI Assistance}}
\newcommand{\categoryEval}{\textbf{Evaluation}}

\definecolor{snapBlue}{cmyk}{0.67,1,0,0.18}
\definecolor{compGreen}{cmyk}{0.92,0,1,0.18}
\definecolor{tempOrange}{cmyk}{0,0.27,0.97,0.24}
\definecolor{tempRed}{cmyk}{0,0.69,0.91,0.32}

\definecolor{myorange}{RGB}{249, 203, 156}
\definecolor{myblue}{RGB}{164, 194, 244}
\definecolor{mygreen}{RGB}{182, 215, 168}
\definecolor{mypurple}{RGB}{180, 167, 214}
\definecolor{myred}{RGB}{234, 153, 153}

\aptLtoXcmd{\newcommand\circlea{\hyperref[activity:a1]{\faCompass}}
\newcommand\circleb{\hyperref[activity:a2]{\faLightbulb}}
\newcommand\circlec{\hyperref[activity:a3]{\faComment}}
\newcommand\circled{\hyperref[activity:a4]{\faHandshake}}
\newcommand\circlee{\hyperref[activity:a5]{\faThermometerHalf}}
\newcommand\allActivities{\circlea{}~\circleb{}~\circlec{}~\circled{}~\circlee{}}
}
{
\newcommand\circlea{\hyperref[activity:a1]{\textcolor{myred}{\faCompass}}}
\newcommand\circleb{\hyperref[activity:a2]{\textcolor{mygreen}{\faLightbulb}}}
\newcommand\circlec{\hyperref[activity:a3]{\textcolor{myblue}{\faComment}}}
\newcommand\circled{\hyperref[activity:a4]{\textcolor{myorange}{\faHandshake}}}
\newcommand\circlee{\hyperref[activity:a5]{\textcolor{mypurple}{\faThermometerHalf}}}
\newcommand\allActivities{\circlea{}~\circleb{}~\circlec{}~\circled{}~\circlee{}}}

\makeatletter
\newcommand\notsotiny{\@setfontsize\notsotiny{6.75}{7.75}}
\makeatother

\newcommand\subscriptsize{\@setfontsize\scriptsize\@viipt{9.5}}

\graphicspath{{figs/}{figures/}{pictures/}{images/}{./}}

\RequirePackage[nameinlink,capitalise]{cleveref}
\crefname{figure}{Figure}{Figures}
\crefname{table}{Tab.}{Tabs.}
\crefname{table}{Table}{Tables}
\crefname{section}{Sec.}{Secs.}
\crefname{section}{Section}{Sections}

\definecolor{vlg}{RGB} {230, 235, 242} 
\newcommand{\opportunity}[1]{
\begin{quote}
\textbf{{Opportunity:}} \textit{#1}
\end{quote}
}

\begin{document}

\title{Challenges in Synchronous \& Remote \\Collaboration Around Visualization}

\author{Matthew Brehmer}
\affiliation{%
  \institution{University of Waterloo}
  \city{Waterloo}
  \state{Ontario}
  \country{Canada}}
\email{mbrehmer@uwaterloo.ca}

\author{Maxime Cordeil}
\affiliation{%
  \institution{The University of Queensland}
  \city{Brisbane}
  \country{Australia}}
\email{m.cordeil@uq.edu.au}

\author{Christophe Hurter}
\affiliation{%
  \institution{ENAC Université de Toulouse}
  \city{Toulouse}
  \country{France}}
\email{christophe.hurter@enac.fr}

\author{Takayuki Itoh}
\affiliation{%
  \institution{Ochanomizu University}
  \city{Tokyo}
  \country{Japan}}

\author{Wolfgang Büschel}
\affiliation{%
  \institution{University of Stuttgart}
  \city{Stuttgart}
  \country{Germany}}

\author{Mahmood Jasim}
\affiliation{%
  \institution{Louisiana State University}
  \city{Baton Rouge}
  \state{Louisiana}
  \country{USA}}

\author{Arnaud Prouzeau}
\affiliation{%
  \institution{Université Paris-Saclay, Inria, CNRS}
  \city{Paris}
  \country{France}}

\author{David Saffo}
\affiliation{%
  \institution{JPMorganChase}
  \city{New York}
  \state{New York}
  \country{USA}}

\author{Lyn Bartram}
\affiliation{%
  \institution{Simon Fraser University}
  \city{Surrey}
  \state{British Columbia}
  \country{Canada}}

\author{Sheelagh Carpendale}
\affiliation{%
  \institution{Simon Fraser University}
  \city{Burnaby}
  \state{British Columbia}
  \country{Canada}}

\author{Chen Zhu-Tian}
\affiliation{%
  \institution{University of Minnesota-Twin Cities}
  \city{Minneapolis}
  \state{Minnesota}
  \country{USA}}

\author{Andrew Cunningham}
\affiliation{%
  \institution{University of South Australia}
  \city{Adelaide}
  \country{Australia}}

\author{Tim Dwyer}
\affiliation{%
  \institution{Monash University}
  \city{Melbourne}
  \state{Victoria}
  \country{Australia}}

\author{Samuel Huron}
\affiliation{%
  \institution{Inst. Polytechnique de Paris}
  \city{Palaiseau}
  \state{Ile de France}
  \country{France}}

\author{Masahiko Itoh}
\affiliation{%
  \institution{Hokkaido Information University}
  \city{Ebetsu}
  \country{Japan}}

\author{Alark Joshi}
\affiliation{%
  \institution{University of San Francisco}
  \city{San Francisco}
  \state{California}
  \country{USA}}

\author{Kiyoshi Kiyokawa}
\affiliation{%
  \institution{Nara Inst. of Science and Technology}
  \city{Ikoma}
  \state{Nara}
  \country{Japan}}

\author{Hideaki Kuzuoka}
\affiliation{%
  \institution{The University of Tokyo}
  \city{Bunkyo-ku}
  \state{Tokyo}
  \country{Japan}}

\author{Bongshin Lee}
\affiliation{%
  \institution{Yonsei University}
  \city{Seoul}
  \country{Republic of Korea}}

\author{Gabriela Molina León}
\affiliation{%
  \institution{Aarhus University}
  \city{Aarhus}
  \country{Denmark}}

\author{Harald Reiterer}
\affiliation{%
  \institution{University of Konstanz}
  \city{Konstanz}
  \country{Germany}}

\author{Bektur Ryskeldiev}
\affiliation{%
  \institution{Mercari R4D}
  \city{Tokyo}
  \country{Japan}}

\author{Jonathan Schwabish}
\affiliation{%
  \institution{Urban Institute}
  \city{McLean}
  \state{Virginia}
  \country{USA}}

\author{Brian A. Smith}
\affiliation{%
  \institution{Columbia University}
  \city{New York}
  \state{New York}
  \country{USA}}

\author{Yasuyuki Sumi}
\affiliation{%
  \institution{Future University Hakodate}
  \city{Hakodate}
  \country{Japan}}

\author{Ryo Suzuki}
\affiliation{%
  \institution{University of Colorado Boulder}
  \city{Boulder}
  \state{Colorado}
  \country{USA}}

\author{Anthony Tang}
\affiliation{%
  \institution{Singapore Management University}
  \city{Singapore}
  \country{Singapore}}

\author{Yalong Yang}
\affiliation{%
  \institution{Georgia Institute of Technology}
  \city{Atlanta}
  \state{Georgia}
  \country{USA}}

\author{Jian Zhao}
\affiliation{%
  \institution{University of Waterloo}
  \city{Waterloo}
  \state{Ontario}
  \country{Canada}}

\renewcommand{\shortauthors}{Brehmer et al.}
\renewcommand{\shorttitle}{Challenges in Synchronous \& Remote Collaboration Around Visualization}

\begin{abstract}
We characterize 16 challenges faced by those investigating and developing remote and synchronous collaborative experiences around visualization. Our work reflects the perspectives and prior research efforts of an international group of 29 experts from across human-computer interaction and visualization sub-communities. The challenges are anchored around five collaborative activities that exhibit a centrality of visualization and multimodal communication. These activities include exploratory data analysis, creative ideation, visualization-rich presentations, joint decision making grounded in data, and real-time data monitoring. The challenges also reflect the changing dynamics of these activities in the face of recent advances in extended reality (XR) and artificial intelligence (AI). As an organizing scheme for future research at the intersection of visualization and computer-supported cooperative work, we align the challenges with a sequence of four sets of research and development activities: technological choices, social factors, AI assistance, and evaluation.
\end{abstract}



\maketitle

\section{Introduction}

\label{sec:intro}

Across educational institutions, businesses, government bodies, and non-profit organizations, people come together to engage in activities that are supported by visualization~\cite{card1999readings}.
These activities include analyzing, generating, monitoring, and presenting data, often with the goal of arriving at data-informed decisions~\cite{brumar2025typology,dimara2021unmet}. 
With broadband internet access now connecting the world, these activities are increasingly being carried out synchronously at a distance, with geographically distributed configurations of participants.
Typically, people use teleconference tools (\eg Microsoft Teams, Zoom, Google Meet) along with a patchwork of productivity tools. 
The global COVID-19 pandemic, with its associated shift to remote and hybrid work, accelerated the use of these tools~\cite{defilippis2022impact}. 
While general-purpose remote collaboration technologies have advanced significantly with respect to multimodal communication via simultaneous video and chat, they do not fully support the nuances of activities that call for synchronous engagement with complex data visualization and analytics artifacts such as computational notebooks and dashboards~\cite{brehmer2021jam}. 
As a result, people manually integrate multiple disjointed tools with limited interoperability, leading to inefficiencies, cognitive overload, and fragmented workflows. 
Consider, for example, a person presenting a slide deck containing static charts, and the audience asks questions that cannot be answered directly with support from the slides, prompting the speaker to toggle to their analytical notebook tool.
These limitations are particularly evident in collaborative data analysis, modeling, and interpretation, where it is critical to seamlessly interact with visualization and analytics artifacts. 
Meanwhile, artificial intelligence (AI) functionality is increasingly prominent in collaboration platforms and visual analytics tools alike, but the social and data context accessible to AI assistants operating in these tools remains siloed.
Finally, consider that in co-located settings, subtle interpersonal cues and nonverbal communication coordinate collaborative visual analysis, where people indicate areas of focus and intent in response to dynamic visualization; maintaining interpersonal awareness is ever more difficult in remote settings.


\aptLtoX{\begin{table*}[tp!]
\caption{
A summary of our challenges across four themes, in order of appearance in (Sections~\ref{sec:challenges:ttt} --- \ref{sec:challenges:evaluation}). 
}
\footnotesize
\centering%
\begin{tabular}{p{1.3cm}p{2cm}p{13cm}}

\cellcolor[HTML]{e9e9e9}\textbf{Theme} &
\cellcolor[HTML]{e9e9e9}\textbf{Index \& Name} &
\cellcolor[HTML]{e9e9e9}\textbf{Challenge}
\\
\midrule
\textsc{\textcolor{black}{techno-}}
&
\ref{sec:challenges:ttt:gap-vis}~
Looking back
& \textit{Reflecting on the viability of approaches initially designed for collaborative visualization across time and space.}
\\
\cellcolor[HTML]{f2f2f2}
\textsc{\textcolor{black}{logical}}
&
\cellcolor[HTML]{f2f2f2}
\ref{sec:challenges:ttt:gap-cscw}~
Looking outward
& \cellcolor[HTML]{f2f2f2}
\textit{Considering the viability of tools, techniques, and technologies initially designed for remote collaborative knowledge work.}
\\
&
\ref{sec:challenges:ttt:asymmetries}~
Asymmetries
& \textit{Anticipating technological asymmetries between collaborators, such as different devices' display and interaction modalities.}
\\
\cellcolor[HTML]{f2f2f2}&
\cellcolor[HTML]{f2f2f2}\ref{sec:challenges:ttt:portability}~
Transferability
& \cellcolor[HTML]{f2f2f2}\textit{Identifying transferable collaborative visualization techniques across scenarios, application domains, and data abstractions.}
\\
\hline
\textsc{\textcolor{black}{social}}
&
\ref{sec:challenges:people:scale}~
Scaling
& \textit{Scaling collaboration activities to accommodate varying numbers of participants, from dozens to hundreds and thousands.}
\\
\cellcolor[HTML]{f2f2f2}
&
\cellcolor[HTML]{f2f2f2}\ref{sec:challenges:people:roles}~
Dynamic roles
& \cellcolor[HTML]{f2f2f2}\textit{Supporting dynamic roles, from contributors to presenters, viewers, and decision makers, with varying levels of expertise and authority.}
\\
&
\ref{sec:challenges:people:agency}~
Agency \& trust
& \textit{Promoting agency and trust as well as awareness of others’ agency, toward a sense of collective ownership and responsibility.}
\\
\cellcolor[HTML]{f2f2f2}&
\cellcolor[HTML]{f2f2f2}\ref{sec:challenges:people:accessibility}~
Accessibility
& \cellcolor[HTML]{f2f2f2}\textit{Increasing accessibility and inclusivity across modalities via direct and indirect approaches.}
\\
\hline
\textsc{\textcolor{black}{ai}}
&
\ref{sec:challenges:ai:paradigms}~
Paradigms
& \textit{Selecting interaction paradigms for AI agents as collaborators, mediators, and assistants, beyond command-and-response exchanges.}
\\
\cellcolor[HTML]{f2f2f2}
&
\cellcolor[HTML]{f2f2f2}\ref{sec:challenges:ai:provenance}~
Provenance
& \cellcolor[HTML]{f2f2f2}\textit{Conveying provenance by capturing and representing the semantics of shared interaction histories across modalities.}
\\
&
\ref{sec:challenges:ai:reliability}~
Reliability
& \textit{Assessing reliability and expectations of AI agents’ capabilities so that human collaborators can assess the risk of error and bias.}
\\
\cellcolor[HTML]{f2f2f2}&
\cellcolor[HTML]{f2f2f2}\ref{sec:challenges:ai:privacy}~
Privacy
& \cellcolor[HTML]{f2f2f2}\textit{Balancing personalization and privacy in shared collaboration spaces, particularly when the former requires extensive data collection.}
\\
\hline
\textsc{\textcolor{black}{evaluation}}
&
\ref{sec:challenges:evaluation:scope}~
Scope
& \textit{Expanding the scope of collaborative visualization evaluation, balancing precision, generalizability, and realism.}
\\
\cellcolor[HTML]{f2f2f2}
&
\cellcolor[HTML]{f2f2f2}\ref{sec:challenges:evaluation:design}~
Questions
& \cellcolor[HTML]{f2f2f2}\textit{From asking the right research questions to selecting appropriate methods and metrics that reflect both individual and group experience.}
\\
& 
\ref{sec:challenges:evaluation:logistics}~
Logistics
& \textit{Navigating the logistics of observing remote collaboration and instrumenting tools to collect multimodal data.}
\\
\cellcolor[HTML]{f2f2f2}
&
\cellcolor[HTML]{f2f2f2}\ref{sec:challenges:evaluation:analysis}~
Analysis
& \cellcolor[HTML]{f2f2f2}\textit{Analyzing richer data capturing remote collaboration; contextualizing across activities and scenarios, and generalizing across groups.}
\\  
\bottomrule
\end{tabular}
\Description{
A Table with three columns and 15 body rows (in addition to a header row). The first column contains the challenge category, the second contains the challenge name and index in the paper, and the third column contains the challenge description.
}
\label{tab:challenges}
\end{table*}}{\begin{table*}[tp!]
\caption{
A summary of our challenges across four themes, in order of appearance in (Sections~\ref{sec:challenges:ttt} --- \ref{sec:challenges:evaluation}). 
}
\Description{
A Table with three columns and 15 body rows (in addition to a header row). The first column contains the challenge category, the second contains the challenge name and index in the paper, and the third column contains the challenge description.
}
\label{tab:challenges}
\notsotiny
\centering%
\begin{tabular}{%
l
l 
l
}

\rowcolor{gray!25}
\textbf{Theme} &
\textbf{Index \& Name} &
\textbf{Challenge}

\\
\midrule
\textsc{\textbf{techno-}}
&
\ref{sec:challenges:ttt:gap-vis}~
Looking back
& \textit{Reflecting on the viability of approaches initially designed for collaborative visualization across time and space.}

\\
\rowcolor{gray!15}
\textsc{\textbf{logical}}
&
\ref{sec:challenges:ttt:gap-cscw}~
Looking outward
& \textit{Considering the viability of tools, techniques, and technologies initially designed for remote collaborative knowledge work.}

\\
&
\ref{sec:challenges:ttt:asymmetries}~
Asymmetries
& \textit{Anticipating technological asymmetries between collaborators, such as different devices' display and interaction modalities.}

\\
\rowcolor{gray!15}
&
\ref{sec:challenges:ttt:portability}~
Transferability
& \textit{Identifying transferable collaborative visualization techniques across scenarios, application domains, and data abstractions.}


\\
\midrule
\textsc{\textbf{social}}
&
\ref{sec:challenges:people:scale}~
Scaling
& \textit{Scaling collaboration activities to accommodate varying numbers of participants, from dozens to hundreds and thousands.}

\\
\rowcolor{gray!15}
&
\ref{sec:challenges:people:roles}~
Dynamic roles
& \textit{Supporting dynamic roles, from contributors to presenters, viewers, and decision makers, with varying levels of expertise and authority.}

\\
&
\ref{sec:challenges:people:agency}~
Agency \& trust
& \textit{Promoting agency and trust as well as awareness of others’ agency, toward a sense of collective ownership and responsibility.}

\\
\rowcolor{gray!15}
&
\ref{sec:challenges:people:accessibility}~
Accessibility
& \textit{Increasing accessibility and inclusivity across modalities via direct and indirect approaches.}

\\
\midrule
\textsc{\textbf{ai}}
&
\ref{sec:challenges:ai:paradigms}~
Paradigms
& \textit{Selecting interaction paradigms for AI agents as collaborators, mediators, and assistants, beyond command-and-response exchanges.}

\\
\rowcolor{gray!15}
&
\ref{sec:challenges:ai:provenance}~
Provenance
& \textit{Conveying provenance by capturing and representing the semantics of shared interaction histories across modalities.}

\\
&
\ref{sec:challenges:ai:reliability}~
Reliability
& \textit{Assessing reliability and expectations of AI agents’ capabilities so that human collaborators can assess the risk of error and bias.}

\\
\rowcolor{gray!15}
&
\ref{sec:challenges:ai:privacy}~
Privacy
& \textit{Balancing personalization and privacy in shared collaboration spaces, particularly when the former requires extensive data collection.}


\\
\midrule
\textsc{\textbf{evaluation}}
&
\ref{sec:challenges:evaluation:scope}~
Scope
& \textit{Expanding the scope of collaborative visualization evaluation, balancing precision, generalizability, and realism.}

\\
\rowcolor{gray!15}
&
\ref{sec:challenges:evaluation:design}~
Questions
& \textit{From asking the right research questions to selecting appropriate methods and metrics that reflect both individual and group experience.}

\\
& 
\ref{sec:challenges:evaluation:logistics}~
Logistics
& \textit{Navigating the logistics of observing remote collaboration and instrumenting tools to collect multimodal data.}

\\
\rowcolor{gray!15}
&
\ref{sec:challenges:evaluation:analysis}~
Analysis
& \textit{Analyzing richer data capturing remote collaboration; contextualizing across activities and scenarios, and generalizing across groups.}

\\  
\bottomrule
\end{tabular}

\end{table*}}


We are by no means the first to reflect on the intersection of visualization and computer-supported cooperative work (CSCW)~\cite{heer2008design,isenberg2011collaborative}.
However, much of the prior work in collaborative visualization either addresses \textit{synchronous} \textit{co-located} collaboration or \textit{asynchronous} collaboration, or it predates the combined rise of immersive social extended reality (XR) technology and real-time AI assistance.
Our goal is thus to focus attention on research and design issues given these current and emerging technologies that will shape how people collaborate around visualization with remote peers. 

This paper contributes a series of \numChallenges~actionable challenges for researchers and practitioners working on synchronous and remote collaboration around visualization (\cref{tab:challenges}); we summarize these challenges in the verb phrases titling the subsections of \cref{sec:challenges:ttt} --- \cref{sec:challenges:evaluation}.
We start with {\categoryTTT} challenges universal to remote and synchronous collaboration around visualization: identifying viable combinations of interface technology and visualization techniques, including recent developments in immersive analytics and XR. 
We then characterize four {\categoryPeople} challenges that may further constrain technical design choices.
Next, given the increasing prominence of AI in collaboration tools and visualization tools alike, we profile the challenges of integrating {\categoryAI} for synchronous and remote collaborative visualization.
Finally, we speak to researchers who study and seek to demonstrate the efficacy of sociotechnical interventions with four challenges relating to {\categoryEval}.
The challenges we pose reflect the perspectives and prior research efforts of 29 authors representing sub-communities across human-computer interaction and visualization, and while they are informed by prior literature, this paper should not be seen as an exhaustive survey.
Throughout the paper, we ground the challenges in five carefully scoped activities (\cref{fig:teaser}): \circlea{} exploratory data analysis,
\circleb{} divergent ideation, \circlec{} presentations around data, \circled{} joint decision making grounded in data, and \circlee{} real-time data monitoring.
For each challenge, we also suggest one or more near-term research opportunities.
An articulation of these challenges will serve to onboard researchers and practitioners to this intersection of topics, and to provide a holistic perspective to industrial partners and funding agencies.


\section{Background: Situating Our Work}
\label{sec:relatedwork}

We structure this paper in a fashion similar to prior work reflecting on the challenges of visualization evaluation~\cite{lam_empirical_2012}, teaching visualization~\cite{bach2023challenges}, and immersive analytics~\cite{ens2019revisiting}.
As in those papers, we are not proposing a new framework or theory, but rather characterizing challenges, scenarios, and research opportunities for the visualization community.

We ground our discussion throughout this paper in categories of activities where conversations are supported by one or more data visualization or visual analytics artifacts (\cref{fig:teaser}). 
These artifacts include individual 2D and 3D representations of data as well those appearing within interactive dashboards~\cite{tory2021finding}, static data reports~\cite{tory2021finding}, slide presentations~\cite{brehmer2021jam}, spreadsheet workbooks~\cite{bartram2021untidy}, and analytical notebooks~\cite{crisan2020passing}.
Across these artifacts, the underlying data could be abstract (i.e., non-spatial) or associated with spatial positions. 
Finally, the data being represented could include historical observations, simulations and future predictions, or outcomes of the activity itself, such as in the case of collaborative ideation.

\bpstart{Intersecting communities, overlapping interests}
We position this work at an intersection of research communities.
Our focus is most aligned with the visualization community, while the collaborative nature of these activities reflects an alignment with the computer-supported cooperative work (CSCW) community. 
While we cast a visualization lens on social- and evaluation-related challenges in this paper, many aspects of them are not unique to visualization applications, so it is therefore helpful to look to CSCW research concerned with other forms of collaborative knowledge work and consider how that research addresses these challenges.
Beyond CSCW, our specific interest in remote and synchronous communication and collaboration around data led us to research territories and emerging technologies being explored by the subcommunities of immersive analytics, social XR, explainable AI, and visualization for communication and education.

An established framework in CSCW~\cite{johansen88} distinguishes activities according to \textit{time} (\ie~synchronous vs.\ asynchronous) and \textit{space} (\ie~co-located vs.\ remote). 
Two often-cited papers reflect on this framework from a visualization perspective: Heer and Agrawala~\cite{heer2008design} present design considerations for asynchronous collaborative visualization taking place in online communities, while Isenberg~\etal~\cite{isenberg2011collaborative} outline challenges and a research agenda spanning the four quadrants of this framework.
In this paper, we are primarily concerned with synchronous and remote (or hybrid) activities.
This quadrant has long been a concern of the CSCW community~\cite{olson2000distance}, however, the COVID-19 pandemic and the pivot to remote work attracted a renewed interest.
Yet, remote collaboration continues to challenge researchers and designers because interpersonal interaction and awareness are difficult to provide without physical proximity~\cite{kraut02}.

While the collaborative visualization guidance offered by Heer and Agrawala~\cite{heer2008design} and Isenberg~\etal~\cite{isenberg2011collaborative} continues to be relevant, we must also acknowledge that it has been eighteen and fifteen years, respectively, since these papers were written.
Their reflections predate the recent rise to prominence of remote knowledge work, AI assistance, and social XR.
These technologies have evolved rapidly in the past half-decade alone, with AI being particularly disruptive. 
Even Ens~\etal's 2021 characterization~\cite{ens2021grand} of challenges in immersive analytics, which includes five challenges relating to collaboration, merits a new reflection on progress made in the face of these developments.
In summary, our present contribution of research challenges should ultimately be viewed as a modernization of these prior discussions of collaborative visualization.

\bpstart{A landscape of commercial collaboration tools}
Today, teleconference tools such as Zoom~\cite{zoom2022}, Webex Meetings~\cite{webex2022}, Google Meet~\cite{meet2022}, and others are omnipresent in workplaces, organizations, and educational institutions. 
These tools typically support real-time multi-party video and audio transmission, screen sharing, breakout rooms, polls, reactions, as well as back-channel and side-channel text conversations. 
These tools are generic enough to address a multiplicity of social interactions, including business meetings, educational lessons, international conferences, and family gatherings.
Often, they are integrated or used in conjunction with cloud-based knowledge work or productivity suites, including collaboration platforms such as Teams~\cite{teams2022} and Slack~\cite{slack2022}, forming synchronous episodes within a larger timeline of asynchronous communication.
Despite the increasing prominence and expanding functionality of these tools, recent studies have shown that co-located collaboration has better cross-brain synchrony compared to screen-based collaboration~\cite{zhao2023separable}, and that remote collaboration tools contribute to what has become known as \textit{Zoom fatigue}~\cite{shoshan2022understanding}.

\bpstart{Toward immersive remote collaboration experiences}
Remote work during the pandemic prompted a collective desire for collaboration experiences that are more immersive than desktop teleconference calls.
This is reflected in workplaces co-opting open-world video games to host meetings~\cite{rockpapershortgun2020} and by the recent resurgence in business travel to pre-pandemic levels~\cite{gbta2024,wttc2024}, incurring both personal and environmental costs.

Extended reality (XR) technology presents another path toward immersive collaboration experiences.
There appears to be renewed willingness to embrace this technology based on the uptick in the adoption of head-mounted displays (HMDs) by enterprise organizations~\cite{si2024}.
Responding to this trend, Microsoft introduced Mesh for Teams~\cite{mesh2025}, and Apple has prioritized collaborative enterprise applications in visionOS, encouraging developers to use its SharePlay framework~\cite{shareplay2025}.
This trend is also reflected in vendors offering bespoke VR office spaces and conference venues for organizations and events (e.g.,~\cite{arthur2025,immersed2025, mootup2025, spatial2025, virbela2025}).

\bpstart{The emergence of AI collaboration assistants}
Finally, collaboration tools are increasingly employing AI assistance, and not only in the form of generated conversation summaries and live transcription.
AI's role is transitioning into an active participant alongside multiple human collaborators~\cite{vaccaro2024combinations}.
Accordingly, one category of challenges we consider in this paper address the impact of AI's expanding role on remote collaboration activities involving shared visualization and analytics artifacts.



\section{Methodology}
\label{sec:methodology}
We began identifying challenges for synchronous and remote collaboration around visualization (\cref{tab:challenges}) in discussions at a 2024 NII Shonan seminar (meeting \#213: \textit{Augmented Multimodal Interaction for Synchronous Presentation, Collaboration, \& Education with Remote Audiences}). 
These discussions continued in weekly remote working sessions with seminar attendees in the following months.  

\bstart{Participants}
Twenty-seven of the authors attended the seminar in person, while an additional two joined remotely.
Our collective expertise spans multiple areas of human-computer interaction research, including information visualization, user interface software and technology, interactive surfaces / spaces, computer-supported cooperative work, and extended reality (or XR: encompassing mixed, augmented, and virtual reality), along with the intersections of these areas (\eg immersive analytics, collaborative visualization, and visualization beyond the desktop).
The authors also brought expertise in several application areas, including education, knowledge work, business intelligence, emergency training / response, and air traffic management. 
Finally, the diversity of the authors is reflected in terms of level of seniority (from postdoctoral researchers to full professors), affiliation type (including academic institutions, industrial research groups, a national research laboratory, and a nonprofit policy research organization), gender, and global representation (\ie~nine countries across four continents). 

\bstart{Seminar focus}
Our seminar was preceded by a half-day workshop at the 2023 IEEE VIS conference~\cite{mercado2023}, which concluded with a panel on remote communication around data. 
Several distinct topics of conversation arose in this panel discussion, including the impact of role diversity in remote communication, the need for situation awareness, the impact of AI and automation, alternative design paradigms, the relative advantages and disadvantages of remote communication relative to co-located communication, and a future research agenda. 
The panel topics seeded the discussion of our 2024 Shonan seminar.
The organizers (\ie the first four authors) also deliberately broadened the scope of the seminar relative to the workshop, to be more inclusive of those with expertise in areas of HCI beyond information visualization. 
To prepare for the event, the organizers introduced several additional topics of discussion: the potential of XR technology for supporting remote communication around data; low-cost technology solutions for supporting the same; the potential of multimodal interaction; and taking inspiration from how professional broadcast journalists communicate narratives supported by visualization to their audiences.

\bstart{Seminar organization}
Our seminar took place over the course of four days, consisting of twelve sessions, each lasting up to two hours. 
After two initial sessions of brief introduction presentations from the organizers and participants, seven of the seminar sessions that followed were breakout group discussions that took place in parallel, while the remaining sessions were dedicated to larger group discussions and presentations from the breakout groups.
The organizers also encouraged attendees to participate in multiple breakout group discussions, and to split or combine breakout groups as their conversations digressed or overlapped.
From these breakout discussions we identified four challenge themes along with an initial set of challenges for each. 

\bstart{Working sessions}
After the seminar, the authors met in weekly virtual working sessions via Zoom between December 2024 and March 2025, each one hour long.
To align with the working hours of contributors across different continents, those in Asia and Australia met every fortnight, while those in Europe and North America met in the interleaving weeks at a different time.    
The first author attended both series of sessions to moderate and relay information between the groups.
Between meetings, the authors used Slack for asynchronous discussion.

\bstart{Artifacts and outcomes} The seminar and working sessions produced several artifacts including digital whiteboards, photographs of physical whiteboards, annotated bibliographies, and slide presentations from breakout groups. 
All of the authors also had access to meeting notes, recordings, and AI summarization for the post-seminar working sessions.

The challenge themes (\cref{tab:challenges}) reflect the final organization of the breakout groups, while the scope of activities and scenarios (\cref{sec:scenarios}) reflects a consensus achieved over the course of the working sessions. 
One to two authors volunteered to lead each breakout group and liaise with the organizing authors. 
Moreover, each breakout group included at least one author who assumed a \textit{top-down} perspective, which involved enumerating possible challenges and opportunities associated the group's topic for each activity. 
Meanwhile, another author assumed a \textit{bottom-up} perspective, which involved collecting prior work relevant to the group topic and reconciling each work's use cases, limitations, and future research with the activities and challenges emanating from the top-down perspective.



\section{Types of Activities Involving Synchronous \& Remote Collaboration Around Visualization}
\label{sec:scenarios}
\begin{figure*}[!h]

\label{tab:scenarios}
\scriptsize%
\centering%
\begin{tabular}{%
c|c|c|c|c
}

\rowcolor{gray!25}
\circlea~\textbf{Exploratory Data Analysis} & 
\circleb~\textbf{Divergent Ideation} & 
\circlec~\textbf{Presentation} &
\circled~\textbf{Decision Making} &
\circlee~\textbf{Data Monitoring}

\\  
$\vcenter{\hbox{\includegraphics[width=0.17\textwidth]{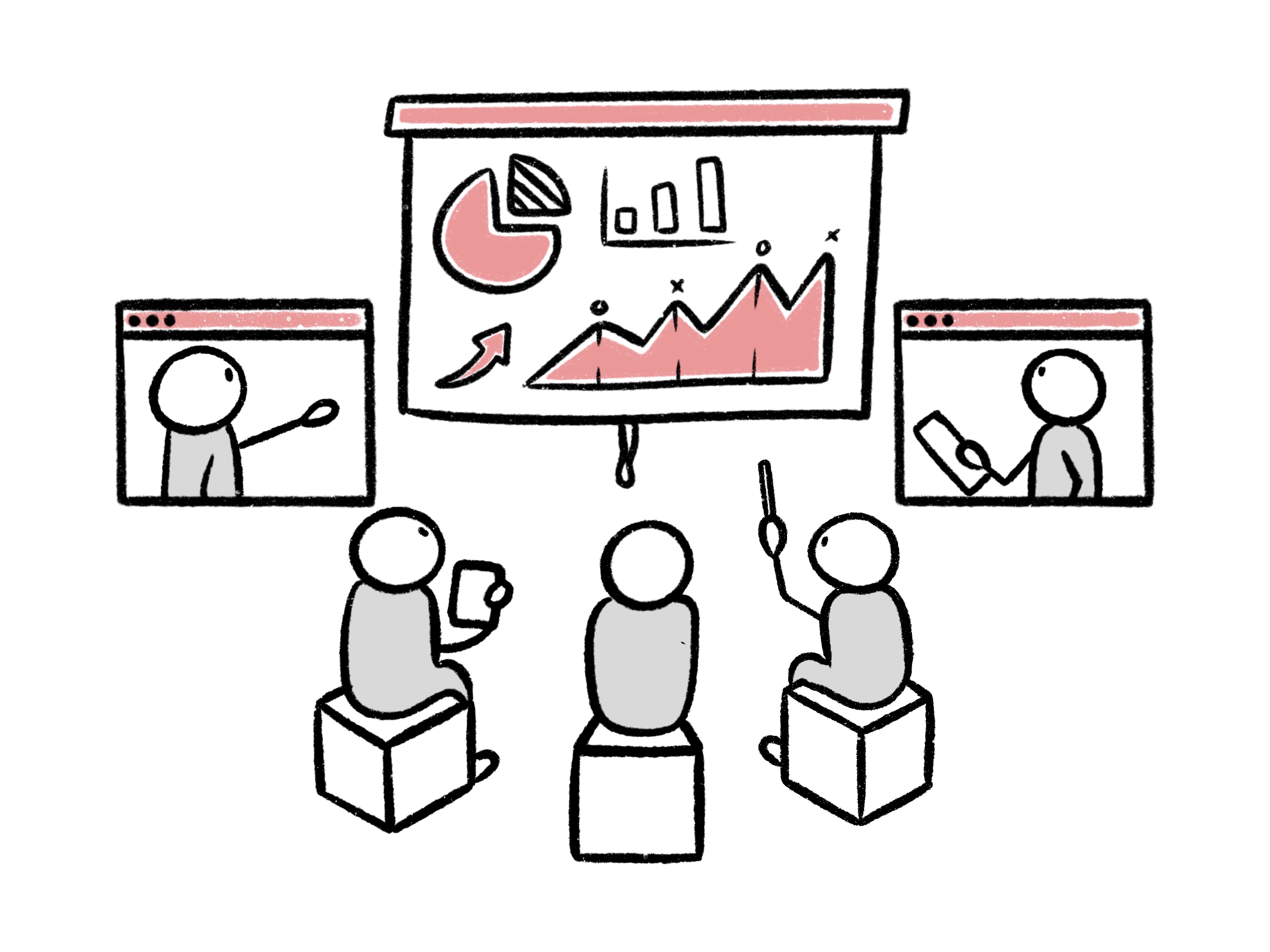}}}$ &
$\vcenter{\hbox{\includegraphics[width=0.17\textwidth]{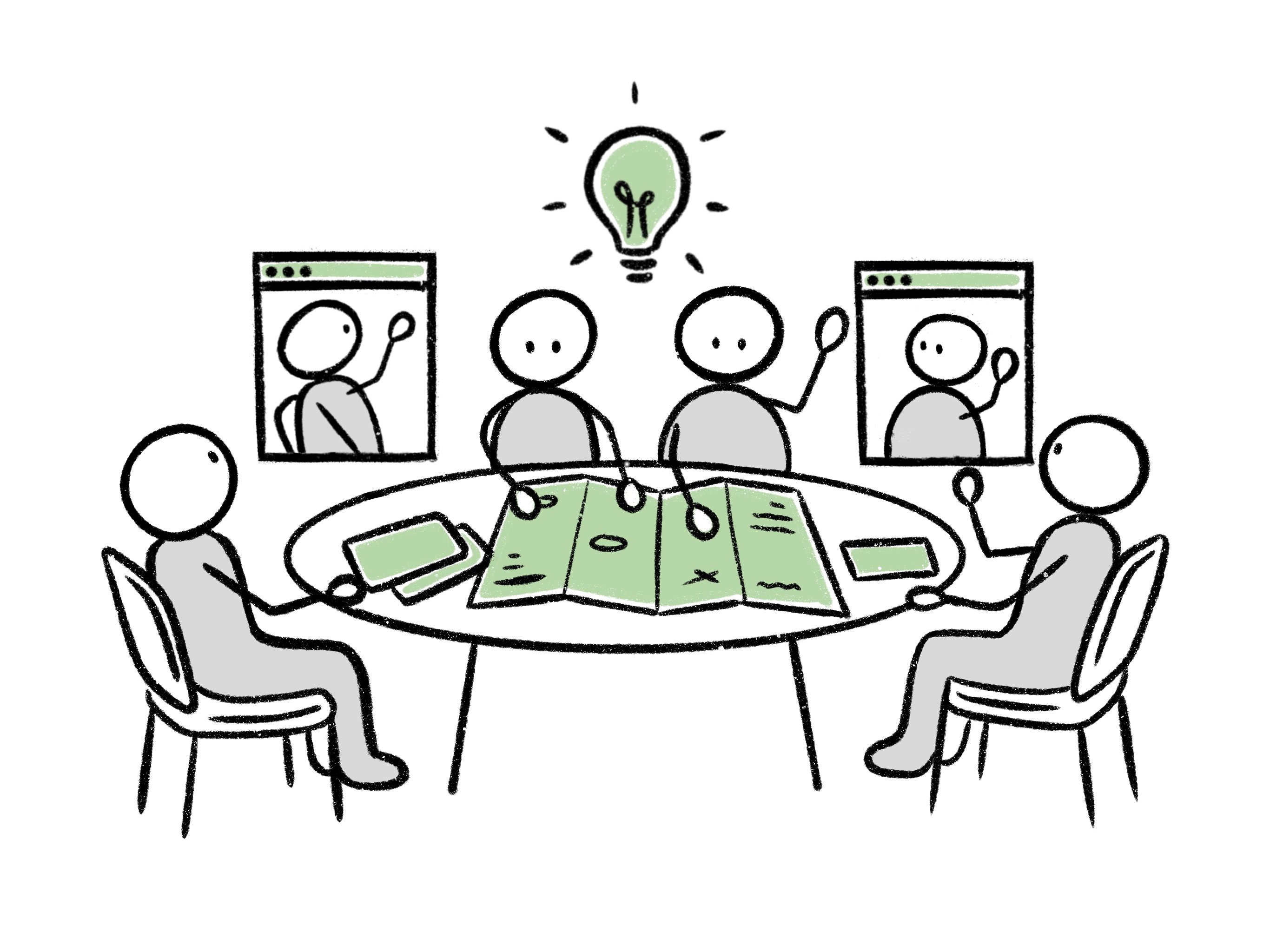}}}$ &
$\vcenter{\hbox{\includegraphics[width=0.17\textwidth]{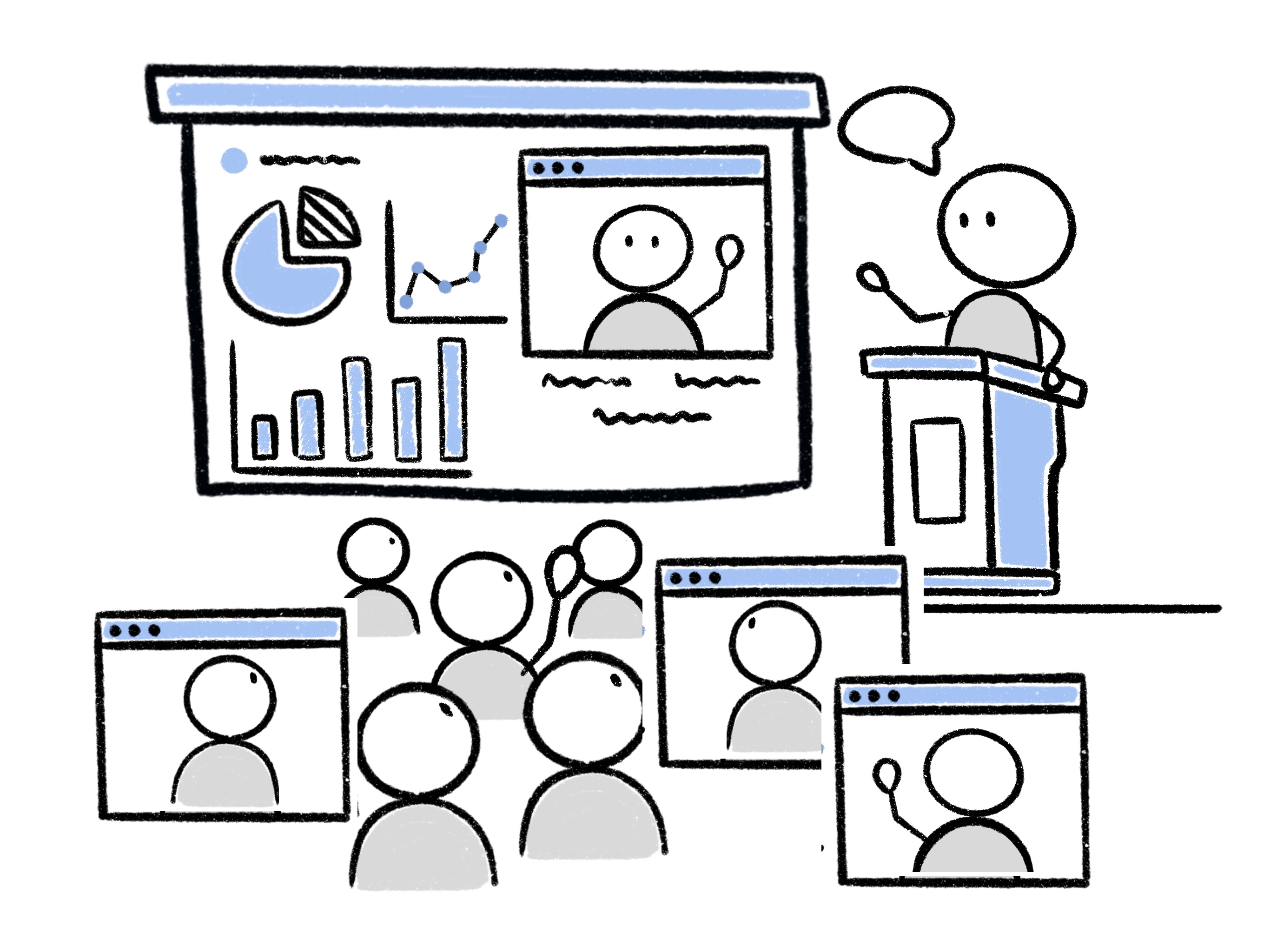}}}$ &
$\vcenter{\hbox{\includegraphics[width=0.17\textwidth]{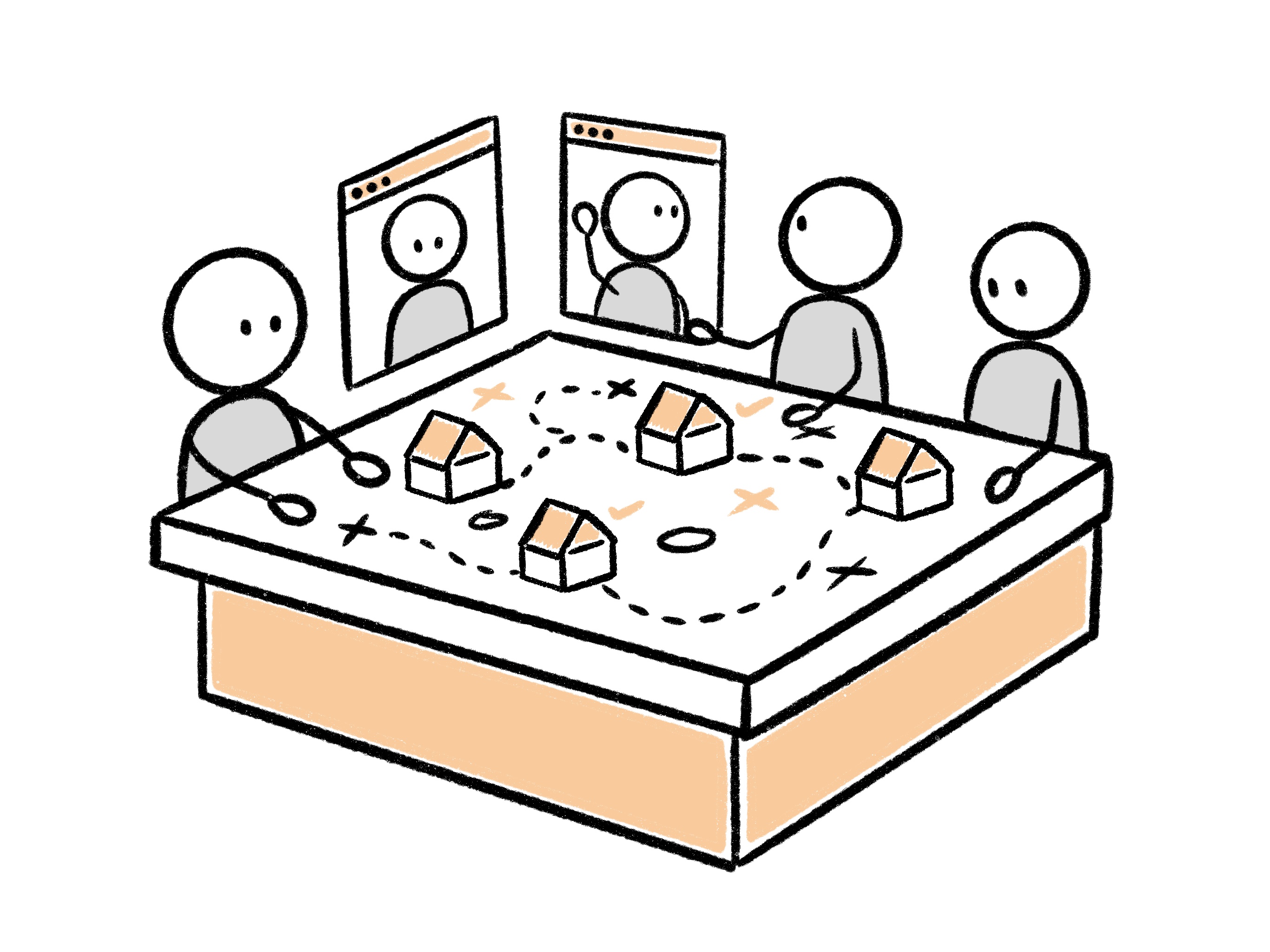}}}$ &
$\vcenter{\hbox{\includegraphics[width=0.17\textwidth]{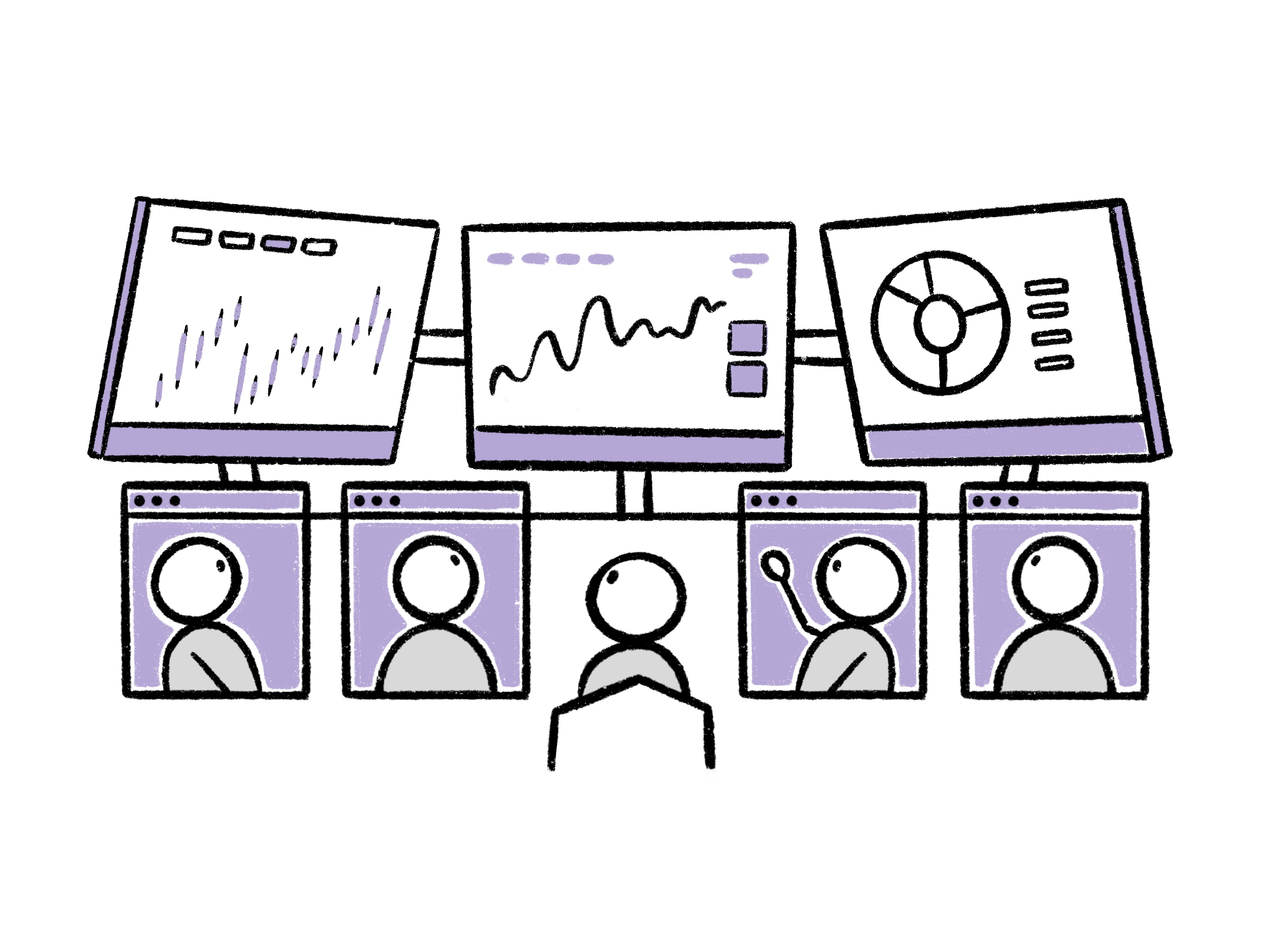}}}$

\\  
\end{tabular}
\caption{
We anchor our challenges for synchronous and remote collaboration around visualization with these five activities. 
}
\Description{Hand-drawn illustrations of five activities involving synchronous and remote collaboration around visualization. In "Exploratory Data Analysis", three local collaborators and two remote collaborators (represented as floating video frames) look at a visualization dashboard together; in "Divergent Ideation", four local collaborators and two remote collaborators appear around a round table with a documents strewn across it, while a light bulb flashes above the table; in "Presentation", a person speaks at a podium next to a projector screen showing a visualization dashboard, while several remote and local audience members watch; in "Decision Making", three local collaborators and two remote collaborators gather around a square table, on which appear representations of houses and roads; in "Data Monitoring", one local collaborator and four remote collaborators look up at three visualization dashboard panels.}
\label{fig:teaser}
\end{figure*}

We anchor our challenges in \textbf{five categories of activity}~(\cref{fig:teaser}, hyperlinked throughout the paper with icons \circlea{} \circleb{} \circlec{} \circled{} \circlee{}). 
Three traits are common to these activities: (1) they involve two or more people; (2) they can take place in co-located, hybrid, or remote settings; and (3) participants use one or more visualization and analytics artifacts.
We adopt this classification because in different activities, collaborators have different goals, and consequently, different needs for the tools they use. 
Our classification also allows us to reason about the specific implications of particular challenges to different activities, which we highlight throughout Sections~\ref{sec:challenges:ttt} ---~\ref{sec:challenges:evaluation}.

\label{activity:a1}
\bpstart{\circlea~Collaborative exploratory data analysis}
We characterize this activity as one where collaborators share a goal of attaining a preliminary understanding of patterns in data~\cite{zhang2020data}. 
While there may be asymmetries of prior knowledge or subject matter expertise among collaborators, their roles are similar in this activity with respect to agency and authority.
This activity may be retrospective (\ie analyzing data collected from the world) or predictive (\ie analyzing data produced by simulation models or generative processes).
Outcomes could include the generation of new hypotheses and ideas, the testing of existing hypotheses, or the identification of new questions or directions of inquiry. 
\begin{quote}
\textit{\textbf{Examples}: business intelligence analysts analyze the sales of product categories across market sectors over time~\cite{brehmer2021jam}; intelligence analysts sift through a document corpus related to a person of interest~\cite{mahyar2014clip}.}
\end{quote}

\label{activity:a2}
\bpstart{\circleb~Collaborative divergent ideation}
Whereas the previous activity described the analysis of collected or modeled data, this one involves a collaborative \textit{generation} of data. 
Despite this difference in data provenance, collaborators may have similar roles and exhibit similar levels of engagement. 
Outcomes could include an enlarged design space or an optimal design given a set of constraints.
In other words, the data generated is the expected outcome of this activity, whereas in the previous one, the expected outcomes are new insights and questions. 
\begin{quote}
\textit{\textbf{Examples}: automotive engineers use a generative process to ideate on building climate-friendly agricultural vehicles~\cite{cvetkovic2023augmented}; interface designers undertake a collaborative co-creation process with AI for designing innovative tangible and embodied interfaces~\cite{shaer2024ai}.}
\end{quote}

\label{activity:a3}
\bpstart{\circlec~Presenting data}
Unlike the previous activities, we characterize presentation as having a marked role and knowledge asymmetry between presenters and audience. There is also an asymmetry of agency, wherein a presenter uses representations of observed or modeled data to support their narrative, directing the audience's attention with multimodal cues, including speech, gesture, and dynamic visual aids~\cite{brehmer2021jam}. 
Meanwhile, the audience could be passive or active depending on the context. 
While audiences in many presentation activities watch passively, more active audiences may question and reply to polls~\cite{huron2013polemictweet} as well as distract or challenge the presenter. 
Ultimately, the intended outcomes of this activity include informing, educating, entertaining, persuading, or provoking the audience, potentially to change what they believe. 
\begin{quote}
\textit{\textbf{Examples}: an instructor of a visualization course introduces students to alternative interaction paradigms~\cite{bach2023challenges}; a public relations team at a publicly traded company reports on earnings to investors~\cite{brehmer2021jam}}
\end{quote}

\label{activity:a4}
\bpstart{\circled~Joint decision-making grounded in data}
This activity presumes that multiple participants are jointly responsible for negotiating and deciding upon a course of action that is informed by observed or modeled data~\cite{dimara2021unmet}, where a course of action could involve choosing among options, agreeing upon threshold values, or creating new artifacts~\cite{brumar2025typology}.
Beyond role and knowledge asymmetries, this activity could involve asymmetries of authority, and the relationships between participants could either be collaborative or adversarial.
Outcomes include achieving a state of consensus or productive dissensus, identifying a compromise, or a resolution to a conflict. 
The collection and visualization of opinions could be performed during the activity to achieve these states~\cite{bressa2024input}.
Like the previous activity, the diversity of relationship dynamics and stakes of the decisions to be made underscore the importance of emotional and nonverbal cues between participants.
\begin{quote}
\textit{\textbf{Examples}: recruiters at a news outlet rank submitted essays to select candidates to interview~\cite{zheng2023competent}; a group of colleagues visiting another country evaluate and decide upon aspects of their itinerary~\cite{han2023redbot}}
\end{quote}

\label{activity:a5}
\bpstart{\circlee~Real-time collaborative data monitoring}
Whereas the previous activities were characterized by the use of observed or modeled / simulated data, a defining characteristic of this activity is the dynamic nature of the data, changing in real time in response to conditions that may or may not be within the collaborators' control. 
In this activity, some of the collaborators may be physically adjacent to the event, system, or process that is generating the data being monitored.
The desired outcome of this activity is to identify urgent threats or deficiencies and to respond accordingly. 
\begin{quote}
\textit{\textbf{Examples}: forecasters and field operators make avalanche risk assessments~\cite{nowak_designing_2024}; coaches, trainers, and athletes coordinate in response to an opposing team's behavior~\cite{gong2024collaborative}.}    
\end{quote}



\section{\underline{\categoryTTT}~Challenges in Synchronous \& Remote Collaboration around Visualization}
\label{sec:challenges:ttt}

As a starting point for supporting synchronous and remote collaboration around visualization, the first four challenges we pose (\cref{tab:challenges}, rows \ref{sec:challenges:ttt:gap-vis}-\ref{sec:challenges:ttt:portability}) pertain to navigating the design space of collaborative techniques and technologies.

\subsection{\underline{Looking Back} on Collaborative Visualization Techniques\ldots}
\label{sec:challenges:ttt:gap-vis}

\bstartnc{\ldots across the space / time framework~\cite{johansen88}.}
Early work, including Sense.us~\cite{heer2008design} and Many Eyes~\cite{viegas2007manyeyes}, focused on \circlea{} analytical activities and \textit{asynchronous} experiences.
Similarly, studies of \textit{co-located} and \textit{synchronous} collaborative data analysis involving pairs~\cite{arias2011pair} or groups~\cite{mahyar2014clip} of people shed light on the interplay between visualization tools and social dynamics, or on how people use shared displays such as tables~\cite{isenberg2011co}.

\opportunity{Survey the affordances of previously proposed techniques in collaborative visualization to assess the viability of each technique for synchronous and remote collaboration.}

Arguably, the advantages of co-located over remote collaboration are weakly defined. 
In \cref{sec:relatedwork}, we remarked upon different brain activation~\cite{zhao2023separable} and fatigue levels~\cite{shoshan2022understanding} induced by screen-based collaboration tools.
However, remote collaboration has benefits beyond personal and environmental costs, with experiences that transcend the conventions of teleconferencing.
For synchronous and remote collaboration around visualization, teleconference platforms' screen sharing functionality has enabled these conversations to take place where one participant \circlec{} presents a dashboard or notebook application~\cite{brehmer2021jam}. However, this limits how other participants can interact with the content, leaving audiences disengaged or feeling left out of \circled{} decision-making processes.
Reminiscent of the real-time collaboration features in Google Docs, recent work~\cite{han2024deixis,neogy2020representing,schwab2020visconnect} has demonstrated analogous collaborative interaction and multi-user awareness with web-based interactive charts, such that participants can see the selections, manipulations, and annotations of others in real time, facilitating the use of deictic utterances when used in conjunction with teleconference tools~\cite{han2024deixis}.
However, these experiences remain unlike those taking place in co-located settings, as participants' interpersonal awareness is split across a patchwork of applications and communication modalities.

\subsection{\underline{Looking Outward} to Tools, Techniques \& Technologies for Synchronous \& Remote Collaboration\ldots}
\label{sec:challenges:ttt:gap-cscw}

\bstartnc{\ldots to assess their viability for supporting visualization activities.}
Some teleconference tools have recently introduced ways to restore interpersonal awareness and non-verbal cues made by a presenter as they screen share content. 
For instance, Zoom and Teams offer functionality to segment the presenter’s outline from their webcam video and composite it in front of screen shared content. 
Virtual camera applications have also become popular, such as OBS Studio~\cite{obs2022} and Airtime~\cite{airtime2025}.
However, only a single presenter can interact with shared content, and they must do so using standard mouse and keyboard interfaces.
The emerging medium of augmented video~\cite{gronbaek2021mirrorblender,liao2022realitytalk} represents a progression from simple speaker segmentation and compositing, integrating gestural control and deictic highlighting to direct an audience's attention, such as in recent demonstrations in \circlec{} data storytelling~\cite{hall2022augmented}.

\opportunity{Survey the affordances of commercial applications and CSCW techniques intended for remote and synchronous collaboration to assess the viability of each technique for collaboration around visualization.}

Departing desktop technology altogether, social XR~\cite{jasche2021beamlite,fink2022relocations} is promising in that it can elicit a feeling among remote collaborators of being in a shared space.
Effective remote collaboration in XR requires precise gesture tracking for effective nonverbal communication, the sharing of proxies for physical artifacts, and support for the awareness of the positions of remote participants' avatars in a common virtual space.
It is therefore unsurprising that synchronous and remote collaboration around visualization also appears among grand challenges in immersive analytics~\cite{ens2021grand}.

\opportunity{Determine if social XR techniques bring participants closer to a feeling of co-located collaboration for activities \allActivities, a feeling of ``bridging the distance gap''~\cite{billinghurst2018collaborative}.}

In 1992, Hollan \& Stornetta~\cite{hollan1992beyond} asked if ``\textit{imitating face-to-face communication} [should] \textit{be the end goal of digital communication.}''
However, 
there is evidence that physical co-presence may not always be beneficial for collaboration~\cite{Born2019colocated}.
The benefits of face-to-face communication (\eg~nonverbal conversation cues) may not be similarly critical between activities \allActivities.
We believe that remote and hybrid collaborative experiences around data can offer something qualitatively different, echoing Hollan \& Stornetta's \textit{`beyond being there'} mentality, one that has continued with recent work discussing the future of XR meetings~\cite{mcveighschultz2022beyond}.
Similarly, Willett et al.'s reflection on `superpowers' as inspiration for visualization~\cite{willett2021perception} includes collaborative capabilities that would be difficult or impractical to realize in co-located settings.
For instance, Willett~\etal proposed \textit{emotion vision} for \circlec{} presentation activities, in which collaborators' video feeds or avatars are augmented with real-time representations of their inferred emotional states.
For visualization-rich presentations, a presenter might find it useful to associate these states with the data observations being shown.

Over the last decade, XR collaboration research has continued to reflect a \textit{`beyond being there'} mentality, such as Kim~\etal's~approach~\cite{Kim2019} which allows participants to inhabit a common visual perspective and augment each other's field of vision with visual cues relevant to the task at hand.
Sharing a perspective in this manner could be particularly beneficial in the presence of 3D visualization artifacts.
Another example is StreamSpace's anachronistic portals that bridge time and space~\cite{ryskeldiev2018streamspace}, in which remote participants can provide situated and perspective-aligned snapshots of a physical environment within a virtual reconstruction of that environment.
This portal approach could be augmented with situated visualization, enabling a real-time collaborative analysis of data associated with a physical environment alongside its digital twin.

Finally, emerging technologies offering new display and interaction modalities may have untapped potential with respect to supporting synchronous remote collaboration 
around data.
Exploring sensory modalities beyond vision and sound (\eg data haptification~\cite{prouzeau2019scaptics}, information olfactation~\cite{batch2020scents}) is also an exciting direction for future work.
Whether these modalities are employed in a complementary or substitutive manner, participants in collaborative activities involving visualization and analytics artifacts stand to gain new levels of interpersonal awareness, or a more acute awareness of the dynamics of changing data in the context of real-time data monitoring \circlee{}.

\opportunity{Expand and document the design space of multisensory interface technology with respect to its applicability to \textit{remote} and \textit{synchronous} collaboration around visualization.}

\subsection{Anticipating Technological \underline{Asymmetries}\ldots}
\label{sec:challenges:ttt:asymmetries}
\bstartnc{\ldots between collaborators.}
When activities \allActivities~manifest synchronously with remote or hybrid arrangements of participants, variations in technological affordances and display capabilities quickly become apparent. 
Similarly, asymmetries in broadband connectivity can also impact the fidelity of the experience.
The impact of these technological asymmetries go beyond simple experiential aspects: device choices will determine how visualization artifacts are perceived and interacted with, particularly when display sizes, resolutions, and stereoscopic capabilities vary, or when input modalities have differing degrees of freedom.

\opportunity{Identify, design, and evaluate instances of remote collaboration around visualization involving asymmetric technologies, and explore the effects of device interoperability on collaborative dynamics and whether activity goals are met.}

From early videoconferencing systems to first-generation XR platforms, there was little flexibility with respect to device interoperability \cite{Steed2012Beaming}, and thus, technological asymmetries were seldom encountered. 
However, with the development of cross-platform middleware, the potential to enable collaboration across asymmetric devices is growing, such as when teams of people interact with the same content via different XR devices~\cite{assaf2024cues}.
Device heterogeneity should not be seen as a drawback, but rather as an opportunity to elicit different perspectives on data.
Moreover, devices range considerably in cost, and this is particularly true of XR devices; we should therefore expect and design for technological asymmetries, wherein a subset of participants use expensive and powerful devices while others use commodity equipment. 
Technological asymmetries also elicit a question of how shared and private content could manifest across display types, 
as prior work~\cite{rijnsburger2017personalized} shows that the personalization of content in collaborative settings can be beneficial.
For example, domain experts can refer to data representations visible only to them during \circled{} decision-making activities, or \circlec{} audiences can answer their own questions without interrupting the presenter. 
Selective asymmetric visibility of data may elicit a greater overall sense of transparency among audiences and provide a sense of immediacy beyond what would otherwise be a passive and indirect consumption experience.

In prior visualization research, technological asymmetries appear in the context of co-located collaborative data analysis~\cite{friedl2024study,horak2019vistribute}, and in scenarios of asynchronous collaboration between a distributed team of analysts~\cite{batch2023wizualization}.
Recently, the DashSpace project~\cite{borowski2025dashspace} within the Spatialstrates collaborative XR platform~\cite{borowski2025spatialstrates} anticipates these asymmetries in synchronous remote or hybrid collaborative data analysis \circlea{}.
Beyond analysis activities, Flow Immersive's \circlec{} data storytelling platform~\cite{dibenigno2021flow} is notable for its accommodation of audiences joining via diverse devices, from desktops to phones and head-mounted displays, in which viewers can assume various perspectives around shared 3D data representations. 
\subsection{Identifying \underline{Transferable} Collaborative Visualization Techniques for Remote Collaboration\ldots}
\label{sec:challenges:ttt:portability}

\bstartnc{\ldots across activities, application domains, and data abstractions.}
One instance of prior applied research in synchronous collaborative visual analytics addresses the well-resourced use case of intelligence analysis~\cite{haeyong2010vizcept}.
As our community accrues visualization design studies that involve remote and synchronous collaboration, we can reflect upon and distinguish between design choices that are uniquely suited for particular application domains from those that may be more easily adapted across domains~\cite{sedlmair2012design}.
Particularly as we consider nascent immersive technologies for remote collaboration around data, we must determine if these technologies are better suited for application domains associated with inherently spatial data (\eg~\cite{mahmood2019improving}) and three-dimensional representations (\eg~~\cite{lee2020shared}) relative to those associated with non-spatial data and a familiar canon of 2D charts and dashboards.

\opportunity{Survey learnings from prior collaborative visualization work conducted in specific application domains and determine aspects that can be adapted across activities \allActivities.}



\section{The \underline{\categoryPeople}~Challenges of Synchronous \& Remote Collaboration Around Visualization}
\label{sec:challenges:people}
In the preceding section, we largely sidestepped discussion about the number of collaborators, the heterogeneity of their roles, and their qualitative experiences when collaborating around visualization, which is the focus of the following four challenges (\cref{tab:challenges}, rows \ref{sec:challenges:people:scale}-\ref{sec:challenges:people:accessibility}).
Overall, these challenges speak to the need for supporting interpersonal awareness and individual differences. While aspects of these challenges are not unique to remote collaboration around visualization, our discussion and the projects we cite in this section reflect our community and the centrality of data in activities \circlea{}~\circleb{}~\circlec{}~\circled{}~\circlee{}.
%
\subsection{\underline{Scaling} Activities \ldots}
\label{sec:challenges:people:scale}

\bstartnc{\ldots to accommodate varying numbers of participants.}
Visualization techniques designed to support synchronous and remote collaboration between pairs of individuals or small groups may not scale to a larger number of participants.
In an investigation of synchronous collaboration and presentation in enterprise business intelligence settings, Brehmer and Kosara~\cite{brehmer2021jam} described activities along the dimensions of scale and formality: (1) small groups of business analysts jointly perform \circlea{} exploratory data analysis, in which a few participants (\ie fewer than ten) share data artifacts and the interactive control of tools; (2) partially structured and partially improvised \circlec{} presentations between cross-functional teams, anticipating interruptions and digressions regarding specific aspects of the data; and (3) \circlec{} formal presentations, such as those prepared for executives, investors, partners, or customers.
Each activity along this progression of scale could feasibly correspond with a 10x increase in the number of participants.
The challenge is therefore to adapt progressions like this across \textit{remote} collaboration.

Part of this challenge is determining the threshold number of participants at which to support particular devices, expose functionality, or prioritize different modalities of communication and interaction. 
For example, immersive analytics research~\cite{yang2022towards} suggests that head-mounted displays (HMDs) could benefit small groups engaged in \circlea{} sensemaking activities, and recent research in astronomical pedagogy~\cite{milisavljevic2025collabxr} demonstrates that classroom presentations can successfully incorporate dozens of HMD-wearing students.
However, it remains unclear if HMDs would remain viable for activities involving \textit{even larger} groups of participants.  
Further complicating this challenge are hybrid collaboration scenarios, in which the magnitudes of co-located and remote participants differ. 
In these cases, designers must strike a balance between leveraging asymmetries while ensuring that co-located and remote participants are similarly engaged.

\opportunity{Classify remote and synchronous collaborative visualization techniques according to how many  collaborators they can accommodate.}

Instances of large-scale collaboration and communication activities in prior visualization research are uncommon, particularly if we are looking for synchronous and remote case studies. 
These activities could scale from dozens of participants~\cite{jasim2021communityclick} to hundreds~\cite{huron2013polemictweet} and thousands or more~\cite{huron2013bubble}.
One such case study is Jasim~\etal's municipal town hall meeting tool that collects and aggregates citizens' feedback as the meeting unfolds~\cite{jasim2021communityclick}, along with their adaptation of this tool for remote town hall meetings~\cite{MERCADOjasim}. 
For events such as these, part of the challenge is determining how to exploit, relay, and summarize both back-channel communication~\cite{dork2010visual,huron2013polemictweet,huron2013bubble} and participants' engagements with shared data artifacts, potentially blossoming into a full-fledged \circlee{} real-time data monitoring or collaborative \circleb{} ideation activity~\cite{bressa2024input}.

\opportunity{Replicate and extend prior studies assessing collaborative visualization techniques with varying numbers of participants.}

When considering the challenge of scale, it can be useful to look beyond remote meetings around data to live broadcasts with remote audiences, particularly for \circlec{} presentation. 
While storytelling and journalism continue to motivate visualization research, a focus on visualization in live television or streaming broadcasts is rare~\cite{drucker2018,huron2013bubble}.
In recent years, it has become increasingly common to see news anchors explain observations in data while interacting with or gesticulating before dynamic representations of data~\cite{cotgreave2020}, whether shown on a large touchscreen display or via a greenscreen composite.
On a smaller scale, data visualization practitioners and educators took to livestreaming platforms during the COVID-19 pandemic to demonstrate and talk about their craft~\cite{cwf2021}.
Both mediums have the potential to engage audiences via back-channel conversations, either as a "second-screen" experience~\cite{neate2015mediating} or as an integrated experience within a livestreaming platform.

\subsection{Supporting \underline{Dynamic Roles}\ldots}
\label{sec:challenges:people:roles}

\bstartnc{\ldots from contributors to presenters, viewers, and decision makers.}
In collaborative knowledge work, activities are often marked by collaborative \textit{phases} and \emph{coupling styles}~\cite{tang06}, referring to the alignment of tasks each person is performing.
Collaboration can also be described as parallel, coordinated, cooperative, or fully collaborative~\cite{castaner2020collaboration}.
Particular categories and combinations of phase and style are often reflected in the affordances of tools and devices, in nonverbal cues for interpersonal awareness, and in participants' territoriality with respect to workspaces and artifacts.
The activities \allActivities~reflect this dynamism of roles.
For instance, in \circlec{} visualization-rich presentations, a participant may transition between being a speaker, a note-taker, a moderator, or an audience member, and in formal presentations about data~\cite{brehmer2021jam}, it is not unusual for one participant to speak while another controls interactive assets like dashboards or slides. 
In other activities, role asymmetries manifest in various ways. 
An \textit{asymmetry of expertise}, whether with respect to the domain or to the tools, changes at different paces for each participant.
In \circled{} decision-making activities in particular, we encounter an \textit{asymmetry of authority}, in which those who lack this authority appeal to or attempt to persuade those who have it.
Meanwhile, in \circlee{} real-time data monitoring activities, we may find an \textit{asymmetry of awareness} along with an \textit{asymmetry of agency}, especially for remote participants responding to or observing an event associated with the data being monitored (\eg emergency response and preparedness, on-the-ground journalistic reporting), wherein the nature of the event may impose situational impairments on these participants.

To respond to the challenge of dynamic roles in remote collaboration, designers should allow participants to componentize data artifacts and surface them on a common application substrate (\eg~\cite{xia2023crosstalk}) where it is possible to grant agency to any participant, in contrast to a single participant screen sharing their local content.
Additionally, in the context of \circlec{} presentation and \circled{} persuasion, componentization could take the form of partially-constructed narratives and alternative argument structures~\cite{kosara2017argument}, allowing participants to adapt their messaging in response to audience engagement.

\opportunity{Design and conduct studies with designated participant role diversity, either by deliberately recruiting those outside of university student participant pools~\cite{friedl2024systematic} and those who self-identify as novices~\cite{burns2023we}, or by assigning roles and providing selective contextual information to different participants prior to and during study execution.} 

\subsection{Promoting \underline{Agency \& Trust} \ldots}
\label{sec:challenges:people:agency}
\bstartnc{\ldots as well as awareness of others' agency.}
Beyond situational impairments, the affordances of tools can bring about an \textit{asymmetry of agency} among participants, particularly if these affordances follow from asymmetries of device and display capabilities (\cref{sec:challenges:ttt:asymmetries}), from scale limitations (\cref{sec:challenges:people:scale}), or from immutable roles (\cref{sec:challenges:people:roles}).
This challenge is concerned with reducing this asymmetry where possible, empowering individuals to make informed decisions and actively contribute to a group activity.
Providing personalized views (\eg\cite{rijnsburger2017personalized}) and other functionality that allows participants to privately and immediately ask their own questions about the data is one way to address this challenge.
Increasing agency also calls for a greater awareness of others' agency, which is especially critical when establishing trust in \circlea{} collaborative data analysis and \circleb{} ideation, in which participants strive toward collective ownership and responsibility with respect to visualization artifacts.
In the context of collaboration activities around visualization, interventions should promote trust both in one's collaborators as well as in shared representations of data.
We revisit the topic of trust below in \cref{sec:challenges:ai} when discussing instances in which these shared representations of data are AI-generated.

\opportunity{Extend recent research on workspace awareness for collaborative productivity tasks in XR~\cite{assaf2024cues} to \circlea{} collaborative data analysis and \circleb{} ideation.}

\subsection{Increasing \underline{Accessibility} \& Inclusivity\ldots}
\label{sec:challenges:people:accessibility}
\bstartnc{\ldots across modalities via direct and indirect approaches.}
Given the multimodal nature of remote collaboration involving shared visualization and analytics artifacts, accessibility issues can limit the engagement of participants with visual, auditory, mobility, or cognitive impairments. 
For example, in the case of vision impairments, data could be represented in alternative ways, including adaptive text descriptions read using text-to-speech assistive tools~\cite{lundgard2021accessible} or displayed via tactile objects~\cite{fan2020constructive}. 
Similarly, tools for \circleb{} collaborative ideation and \circlec{} presentation authoring can integrate established accessibility tools and workflows~\cite{das2019doesn}. 

Beyond accessible representations of data and collaborator activity, this challenge also calls for inclusive interaction design. 
Prior work distinguishes between two inclusive design approaches: \textit{indirect} approaches such as creating or adopting existing accessible interaction design guidelines~\cite{strantz2021using}; and \textit{direct} approaches, or developing ways to modify a tool's affordances to adhere to accessibility requirements, such as by using a screen reader or a sensory substitution tool to interactively navigate data representations (\eg~\cite{chundury2023tactualplot}). 

Inclusive design for collaboration tools also encompasses connecting multilingual and multicultural sets of collaborators, as well as those who lack context. 
Summarization, transcription, and activity replay tools for asynchronous reviewing speak to these needs, however in the case of activities \allActivities, care must be taken to capture the state of shared visualization and analytics artifacts alongside recorded conversations.

\opportunity{Survey synchronous and remote collaborative visualization techniques using Kim~\etal's comparative approach to assessing visualization accessibility~\cite{kim2023beyond}.}



\section{Challenges of Integrating \underline{\categoryAI} into Synchronous \& Remote Collaborative Around Visualization}
\label{sec:challenges:ai}

Artificial intelligence (AI) assistance is increasingly prevalent in commercial communication and collaboration applications, such as Zoom's~\cite{zoom2022} live transcription and meeting summarization features.
Anticipating a further expansion of AI functionality, HCI research has recently explored the potential of proactive AI assistance for moderating meetings~\cite{chen2025we}, onboarding newcomers to existing conversations~\cite{shin2023introbot}, and encouraging ideation~\cite{shaer2024ai}.
Meanwhile, visualization research has also been investigating the integration of AI assistance in the form of conversational question-answering interfaces in visual analytics tools~\cite{hoque2022chart}.
The challenges in this section therefore address the inevitable integration of this AI assistance in synchronous and remote collaboration around visualization.

\subsection{Selecting Appropriate Interaction \underline{Paradigms}\ldots}
\label{sec:challenges:ai:paradigms}

\bstartnc{\ldots for AI agents as collaborators, mediators, and assistants.}
AI's effectiveness as an assistant, collaborator, or mediator depends on its alignment with a group's expectations and trust levels, as well the goal of the activity.
When engaged in technology-mediated collaborative activities, people build working relationships and a shared mental model of their collaborators through interaction~\cite{gilson2015virtual}. 
Despite the benefits of AI assistance to enhance remote collaboration~\cite{zheng2023competent, shin2023introbot}, existing interaction models rely heavily on natural language communication~\cite{shaer2024ai, gao2024collabcoder, lee2024conversational}, which may not sufficiently capture the broader context that human collaborators have, context apparent through implicit cues and procedural task knowledge that is not externalized in language~\cite{zheng2023competent, liao2020questioning}; instead, this task knowledge may be externalized in visualization artifacts.
This disparity leads to misunderstandings and unpredictable AI responses that hinder collaboration, especially in settings where people with varying levels of experience with respect to AI collaborate together~\cite{zheng2023competent}. 
Furthermore, AI agents are often seen as \textit{``black boxes''}~\cite{dwivedi2023explainable} and are expected to serve as informational assistants rather than as productive collaborators.
These perceptions constrain interactions to command-and-response exchanges that in turn limit the ways in which human collaborators can clarify their intentions. 

As AI models grow more sophisticated~\cite{liang2022advances}, it is crucial to develop systems that both \textit{support} and \textit{understand} human collaborators~\cite{liao2020questioning, arrieta2020explainable}. 
When collaborating around visualization artifacts, these models must integrate these artifacts into their context.
Without such advancements, human-AI miscommunication risks persist, leading to inefficiencies, frustration, and reduced collaborative effectiveness~\cite{zheng2023competent, gao2024collabcoder}, particularly in the case of synchronous and remote collaborative activities where miscommunication \textit{between people} is a problem enough on its own.
This challenge calls for new interaction and feedback paradigms to mitigate these risks, from explaining feedback in XR environments~\cite{xu2023xair} to explicitly assigning new roles to AI agents beyond information retrieval assistants.
These roles might involve taking initiative as a fellow qualitative \circlea{} data analyst~\cite{gao2023coaicoder}, recommending visualization aids to a \circlec{} presenter based on audience profiles and questions asked~\cite{takahira2025visaider}, facilitating and eliciting ideas from human collaborators in \circleb{} ideation activities~\cite{ganji2018ease,shaer2024ai}, and moderating deliberations in \circled{} group decision-making~\cite{gao2024collabcoder}. 
Each of these roles could potentially be reinforced by giving these agents a multimodal embodied presence as an avatar~\cite{schmidt2020intelligent}.
To this end, carefully designed personification through avatars could enable more natural and empathetic interactions during human-AI collaborations.

\opportunity{Compare collaborative visualization interfaces in which an AI assistant either: (a) passively monitors conversations unless explicitly invoked; (b) offers side-channel contextual commentary on shared visualization artifacts; or (c) proactively modifies visualization artifacts.}

\subsection{Conveying Interaction and Analytic \underline{Provenance}\ldots}
\label{sec:challenges:ai:provenance}
\bstartnc{\ldots by capturing and representing relevant shared interaction history.}
As collaboration assistants, AI agents can promote a sense of continuity~\cite{drouhard2017aeonium, gao2024collabcoder, yeo2024help,chen2025we,shaer2024ai}.
However, without a clear understanding of how or why attempts to maintain continuity are generated, they may lead to misinterpretation~\cite{zheng2023competent}, a loss of trust~\cite{shaer2024ai}, or dismissal by human collaborators~\cite{gao2024collabcoder}. 
Prior work demonstrates how AI agents visualize past interactions with human collaborators~\cite{buschel2021miria, narechania2021lumos, isaacs2014footprints}.
This work also highlights the critical need to understand the semantics of these interactions.
Without these semantics, we limit the opportunities for manipulating visualization artifacts or iterating upon the AI agents' interpretations. 

\opportunity{Adapt visualization research platforms (\eg~reVISit~\cite{nobre2021revisit}) that support the recording and analysis of non-collaborative interaction with visualization applications, allowing researchers to record and distinguish interactions with collaborative visualization applications accessed synchronously by multiple actors, including AI assistants.} 

As interactions between remote human collaborators, visualization artifacts, and AI agents become increasingly complex in collaborative activities~\cite{wang2019gesture, ou2003gestural, buschel2021miria}, it is critical to develop approaches to effectively capture, classify, and visualize human collaborators' interactions across multiple modalities, such as where they point and click, the text they type, the gestures they make, and the things they say. 
In the case of collaborative immersive analytics, an even richer interaction history is needed, one incorporating the topology of participants and artifacts, such that each participant's intent could be modeled via their proxemic and embodied interaction~\cite{buschel2021miria} with these artifacts.
Without robust representations of semantics in this interaction history, AI systems may not be able to fully make use of previous actions and insights~\cite{yan2024human, gao2023coaicoder}. 
This lack of continuity can result in disjointed communication~\cite{lee2024conversational}, inefficient task progression~\cite{fan2022human}, and diminished trust~\cite{zheng2023competent} in AI's role.
Future AI agents for collaboration around visualization must consider activities beyond \circlea{} collaborative data analysis, capturing and visualizing the provenance of \circleb{} collaborative ideation across multiple interaction modalities and communication channels~\cite{wang2019gesture, ou2003gestural}, or by highlighting and attributing both human and AI contributions to deliberations in \circled{} collaborative decision-making~\cite{shaer2024ai, hubenschmid2022relive}.

\opportunity{Explore and compare approaches that communicate provenance information synchronously and those that surface it retrospectively after synchronous activities end.}
\subsection{Assessing \underline{Reliability}\ldots}
\label{sec:challenges:ai:reliability}

\bstartnc{\ldots and aligning expectations of AI agents' capabilities when collaborating around visualization}.
Contemporary AI agents demonstrate a potential to support both creative and analytical tasks, however when initiated into collaborative activities, the expectations that human collaborators have with respect to an agent's capabilities and its reliability may not consistently align. 
AI agents can only learn from the training data provided, data that may contain biases and lack relevant information about the collaboration context. 
Prior work suggests that placing trust in an AI agent as it completes a task correlates with the tendency to delegate the task \cite{lubar2019ai}. 
Trust is also fragile; high expectations or minor errors can quickly deteriorate it \cite{gao2023coaicoder}, making it challenging to use AI-generated content in a synchronous collaborative activity. 
Although demonstrations of generating and interacting with visualizations via natural language~\cite{narechania2021nl4dv, leon2024talk} predates the rise of large language models (LLM), the advent of these models has led to increasingly complex visualization and explanation~\cite{shen2024ai,tian2025ai}.
Capable of processing complex prompts (\eg those using images as input), these LLMs provide new modalities through which to provide context. 
However, these same capabilities add complexity when assessing their reliability.
Moreover, the artifacts produced by these models may propagate errors and biases that negatively affect how people interpret them~\cite{kong2018ai}. 

\opportunity{In \circleb{} collaborative ideation activities, assess techniques that allow human collaborators to distinguish between human- and AI-generated content.}

Improving the trustworthiness of AI agents is one of the most significant challenges in AI research today~\cite{aaaipanel2025}. 
Determining whether these agents are trustworthy in synchronous collaborative activities requires continuous vigilance on the part of human collaborators; given the non-deterministic behavior of LLMs, it is imperative to provide reliability metrics that help human collaborators assess model outputs together.

\opportunity{Assess dedicated backchannels for fact-checking and bias mitigation of AI-generated visualization artifacts shown in \circlec{} presentations, supporting \circlee{} real-time collaborative monitoring among audience members.}

To successfully integrate AI agents in activities \allActivities, we must identify the tasks that can be safely delegated or supported by these agents. 
Recent work investigating the visualization literacy of LLMs \cite{hong25ai} suggests the possibility of agents acting as \circlea{} fellow visual analysts who can visually annotate features of interest, provided that human collaborators can tune and direct this capability.
However, we require methods of assessing the reliability of this behavior, particularly as models evolve or supersede one another. 
To this end, recent results~\cite{sharma2025can} are promising with respect to using LLMs to understanding spoken conversations about charts, suggesting the viability of using AI assistance to annotate and manipulate charts as a conversation progresses in real time.
 
\opportunity{In \circled{} collaborative decision-making activities, assess how to clearly identify AI agents' modifications of shared visualization artifacts, so as to provide human collaborators relevant metadata so that they can assess risk of error or bias.}
\subsection{Balancing \underline{Personalization} \& Privacy\ldots}
\label{sec:challenges:ai:privacy}
\bstartnc{\ldots in shared collaboration spaces.}
AI-assistance in collaborative activities present opportunities for enhanced customization and personalization at the individual and group level, adapting dynamically to human collaborators’ preferences and interaction patterns. 
Prior research suggests that personalization enhances efficiency and engagement in remote \circleb{} collaborative ideation activities~\cite{xia2023crosstalk, liu2023visual}. 
However, this personalization necessitates extensive data collection, surfacing ethical concerns with respect to data security and visibility when collaborating around visualization. 

\opportunity{Design and evaluate adaptive~\cite{yanez2025state} or adaptable visualization and analytics techniques for use in synchronous and remote collaboration activities, so as to support individual differences~\cite{liu2020survey}.}

Historically, privacy considerations in remote collaboration have revolved around permission settings, such as toggles for camera and microphone access. 
With the advancement of AI capabilities, systems are increasingly requesting a greater diversity of data, encompassing environmental information (\eg~objects and their spatial arrangements in collaborators' surroundings), spoken utterances, facial expressions, and documents related to the topic of the conversation, such as emails, search histories, and calendars. 
This broadening of data collection complicates the definition of privacy boundaries, where vulnerabilities could lead to data misuse, unauthorized access, and a loss of trust. 
An erosion of trust is likely to lead to a rejection of AI agents in \circlea{} collaborative data analysis and \circled{} decision-making, particularly in application domains like finance, healthcare, and education, where sensitive data protection is paramount. 
To address this challenge, tool builders must integrate privacy-preserving methodologies, such as differential privacy, federated learning, and on-device processing, each of which minimizes the exposure and centralization of collaborators' data. 
For instance, whenever possible, personal sensitive data should be processed on-device or at the network edge, so raw information never leaves a person’s local environment~\cite{kleppmann2019local}. 
In other words, personalization strategies should not depend on continuous surveillance or the permanent storage of sensitive information. 

\opportunity{Design and evaluate collaborative visualization techniques that responsibly collect and ensure the privacy of the personal data used to generate personalized interfaces.}



\section{Challenges in the \underline{\categoryEval} of Synchronous \& Remote Collaboration Around Visualization} 
\label{sec:challenges:evaluation}
As collaborative visualization activities are rapidly evolving, such as by adopting AI assistance techniques and social XR technologies, evaluating the design and use of new technologies and techniques to support these activities must necessarily address both \textit{formative} and \textit{summative} questions.
Formative assessment happens at the stages of characterizing the problems and improving design, while summative studies examine how well a given design affects performance and usability.
The visualization research community, and particularly its information visualization sub-community~\cite{lam_empirical_2012}, has traditionally relied on the scientific method of controlled experiments as summative evaluation, a method yielding statistically validated results. 
Defining research questions with testable hypotheses is accordingly a common practice. 
While these methods are highly rigorous and reproducible, they may not fully reflect the richness and complexity of the ``real world''~\cite{sedlmair_information_2011}, inviting questions of ecological validity.  
While these methods form an essential part of an evaluation timeline, researchers must complement them with formative studies that seek to understand practices and issues in the field~\cite{antunes2008structuring}.       
Moreover, unlike task-based visualization studies with individual participants, evaluating collaborative visualization experiences introduces the distinct research and practical challenges discussed in this section.

\subsection{Expanding the \underline{Scope}\ldots}
\label{sec:challenges:evaluation:scope}
\bstartnc{\ldots of collaborative visualization evaluation.}
While the visualization research literature contains many examples of evaluations focusing on individual task performance~\cite{6634108}, there are fewer instances of evaluations examining collaboration around data (\eg~\cite{brehmer2021jam,isenberg2011collaborative}).
In these evaluations, our methodologies must account for a range of possible group dynamics, where hierarchies, personalities, and mixed social and professional backgrounds interact and ultimately determine a group's performance~\cite{antunes2008structuring}. 

\opportunity{Conduct ethnographic inquiries, multi-dimensional in-depth long-term case studies~\cite{shneiderman2006strategies} and longitudinal insight-based evaluations~\cite{saraiya2006insight} across application domains to better understand remote and hybrid work environments and their collaborative work practices involving visualization artifacts.}

As remarked upon in \cref{sec:relatedwork}, much of the prior work in collaborative visualization addresses asynchronous or co-located synchronous collaboration and \circlea{} analysis activities, and the evaluations reported in this work are summative in nature, focusing on usability and lower-level task performance.  
Studies focusing on collaborative activities \allActivities{} and the needs of systems to support them are only recently emerging~\cite{oral2023information}. 
Lam~\etal's questions for evaluating visual data analysis and reasoning are well-aligned with the activities of \circlea{} exploration, \circled{} decision-making, and \circlee{} monitoring.
However, new research questions and data collection methods may be needed for \circlec{} presentation and \circleb{} ideation activities. 
The former recalls Lam~\etal's scenario of evaluating communication through visualization where questions relate to how an audience learns from visualization artifacts and how a visualization tool helps people explain or communicate concepts.
 
\opportunity{For \circlec{} presentation activities, conduct post-presentation interviews and focus groups with audiences.}

Beyond expanding the scope of evaluation to consider more synchronous and remote activities around visualization, there are also opportunities for future studies to investigate the impact of the emerging technologies discussed in this paper, namely AI assistance and XR.

When planning any evaluation, we seek \emph{precision}, \emph{generalizability}, and \emph{realism}, but achieving all three is rare; we must make compromises and innovate with respect to how we capture salient details of synchronous and remote collaboration around visualization. 
If prioritizing precision, we could strive to design simulations (\eg~\cite{tong2023towards,enriquez2024evaluating}) of activities \allActivities{}, with assigned participant roles, prescribed tasks, and multimodal communication channels instrumented to capture group dynamics. 
However, the assigned roles and low stakes of simulated activities are unlikely to expose the effects of domain expertise and collaborator rapport on distributed cognition in a \circlea{} collaborative analysis session. 
Nor are they likely to elicit genuine engagement and interest in the topic of a \circlec{} presentation.
Finally, they are unlikely to produce a pressure to produce results that underlie the give-and-take of \circleb{} group ideation or the negotiation process in \circled{} joint decision-making.

\subsection{From Asking the Right Research \underline{Questions}\ldots}
\label{sec:challenges:evaluation:design}
\bstartnc{\ldots to selecting appropriate methods and metrics.}
Studies of synchronous and remote collaboration around visualization can have different goals.
For those studying existing practices, a goal could be to inform the design of future systems, while for those who have built a novel system or technique, goals might include evaluating its utility or usability. 
The goals in turn determine research questions and methodological choices, including the definition of metrics that reflect both individual and group-level experience, as well as the operationalization of how these metrics are measured and analyzed. 

Research questions are further shaped by the logistical complexities of studying groups of varying size, structure, and role heterogeneity~\cite{neale_evaluating_2004}.
As researchers undertake a series of studies reflecting an iterative process of requirements analysis, design, and summative evaluation, asking the right research questions at the outset of this process is critical.
A diversity of research questions demands a more comprehensive toolbox that includes (among others) empirical, observational, participatory, co-design, and reflective methods~\cite{neale_evaluating_2004,lam_empirical_2012,hogan2015elicitation}. 
There is a perception that more qualitative approaches are not sufficiently verifiable and lack external validity (\ie~generalizability)~\cite{6634108}, although they are common to critical systems where collaborative decision-making is essential~\cite{woods2019observations,vivacqua2016collaboration}. 
In the visualization community, experience with these methods, and how to refine the research questions that benefit from them, is beginning to emerge, and we must be willing to employ these methods in the study of collaboration around visualization.

\opportunity{Design and conduct studies that consider multiple aspects of collaboration around visualization, including effectiveness and efficiency, group insight generation, and group dynamics including engagement and the facilitation of social exchange.}

Framing the problem space for collaborative visualization activities beyond analysis is more likely to generate new research questions than answer existing ones.  
Some prior work has taken steps in this direction, including a study of \circled{} decision makers within organizations~\cite{dimara2021unmet} and a study of experts who routinely engage in \circlee{} collaborative distributed data monitoring~\cite{nowak_designing_2024}.
While neither of these studies resorted to complete ethnographic immersion, defining what constitutes a \circleb{} “good” but divergent idea or a \circled{} “good” decision may necessitate a more ethnographic approach to fully understand possible nuances.

\opportunity{Survey research questions and methods used to study collaborative systems and particularly collaborative visualization systems intended for remote and synchronous collaboration.}


\subsection{Navigating the \underline{Logistics}\ldots}
\label{sec:challenges:evaluation:logistics}

\bstartnc{\ldots of observing synchronous and remote collaboration around visualization.}
Observing the use of complex systems involving asymmetries of role and technology can be arduous~\cite{reski2022empirical,tong2023towards}. 
Unsurprisingly, constrained by technology, many prior studies focus on unimodal interaction and symmetric collaboration~\cite{isenberg2011collaborative,olson2000distance}. 

Looking to the future, researchers will need to coordinate with multiple participants and instrument various devices and applications to collect observations across locations~\cite{wall2022vishikers}.
This will require careful scheduling and access to tools and devices, access that is complicated by remote participants' use of their own devices, from computers to mobile devices and head-mounted displays (HMDs). 

\opportunity{Adapt interaction logging tools (\eg~reVISit~\cite{nobre2021revisit}) for synchronous and remote collaborative visualization activities spanning multiple devices.}

Throughout this paper, we have emphasized the role of multimodal communication across collaborative activities \allActivities{}, which suggests a need to collect observations across the modalities~\cite{koch2025multimodal} of speech, gesture, gaze, and touch. 
However, subtle communication cues can easily be lost in remote collaboration settings without careful instrumentation.
A recent review~\cite{friedl2024systematic} shows that such multimodal instrumentation is already apparent in the evaluation of immersive analytics systems, suggesting that we could similarly instrument tools and devices for remote collaboration around data, and not just for activities involving immersive technology.
We stress the importance of multimodal data collection because familiar evaluation methods such as the ``think aloud protocol'' are impractical in collaborative activities. 
Finally, multimodal data collection also invites additional scrutiny with respect to data security and privacy, for combinations of observations across modalities can be used to de-anonymize participants or reveal sensitive material.

Following from the challenge of scale (\cref{sec:challenges:people:scale}), special logistical consideration is warranted for large-scale collaborative activities, such as those with hundreds or thousands of participants and a greater diversity of roles~\cite{dork2010visual,huron2013polemictweet}.
For such activities, evaluation is often opportunistic, for which we can study a targeted deployment of a prototype tool that allows people to \circlee{} monitor and discuss real-time data associated with a significant event, such as an election, a public forum, a climate emergency, a sporting event, or a cultural spectacle.
Large-scale \circlec{} formal presentations focusing on data~\cite{brehmer2021jam} where audiences interact with one another via backchannel conversations are particularly interesting instances to consider.
While prior ``in the field'' studies have shed light on large-scale, ecologically-valid, and collaboration activities~\cite{jasim2021communityclick,huron2013bubble,huron2013polemictweet}, these studies impart additional complexity, such as by coordinating with a cast of stakeholders that could impose legal, technical, or organizational constraints on research. For example, experimentation during a live TV show~\cite{huron2013bubble} or a conference~\cite{huron2013polemictweet} could involve security delays, as stakeholders bear legal responsibility for broadcast content.

\subsection{\underline{Analyzing} Richer Data\ldots}
\label{sec:challenges:evaluation:analysis}
\bstartnc{\ldots capturing synchronous and remote collaboration around visualization.}
Applying qualitative, quantitative, or mixed-method frameworks to the analysis of collaborative activities can each provide distinct insights.

Researchers evaluating collaborative technology commonly collect both quantitative and qualitative data, combining interaction logs, surveys, and questionnaires with video recordings and ethnographic observations~\cite{Pinelle2000}. 
In doing so, they capture task performance as well as subtler aspects of group dynamics, such as workspace awareness~\cite{Gutwin2002} or social presence~\cite{Biocca2001}, constructs that are pivotal to \circlea{} collaborative analysis and \circleb{} ideation activities.
However, quantitative observations can overlook nuanced interpersonal cues critical to effective collaboration, and standardized questionnaires are sometimes too generic to capture context-specific phenomena~\cite{neale_evaluating_2004}. 
Similarly, interaction log data may mask small but meaningful signals of coordination spread across modalities and timeframes~\cite{Yang2021}.
Meanwhile, video analysis and ethnographic methods can be time-consuming, are dependent on identifying the right unit of analysis, and are prone to observer bias or the tendency to focus on specific details~\cite{Plowman1995}. 
Moreover, in high-stakes \circled{} decision making activities, these methods may be intrusive absent an established rapport with participants.
As a consequence, the volume and diversity of observations can overwhelm traditional analysis approaches, making it difficult to identify overarching patterns.

Ultimately, researchers must construct meaning from their observations. 
When studies yield both observations of individual participants as well as observations of group dynamics, merging these perspectives does not guarantee a comprehensive understanding of the larger collaborative processes at play. 

\opportunity{Cohesively interpret findings from studies of synchronous and remote collaboration around visualization and contextualize them within the fabric of ongoing longitudinal asynchronous collaboration.}

Finally, groups will differ in size, composition of co-located and remote collaborators, distribution of expertise, and device usage, and while these factors will make it difficult to generalize findings, a thorough accounting of the idiosyncrasies of group dynamics will provide useful nuance to other data collected.



\section{Final Reflections}
\label{sec:discussion}
In this paper, we contributed a characterization of \numChallenges~challenges pertaining to synchronous and remote collaboration around visualization that present entry points for a range of stakeholders.
For \textbf{researchers}, our work presents an organizing scheme for future work, with one or more near-term research opportunities suggested for each challenge.
While any of the other challenges identified in this paper could become the focus of enterprising young researchers and students, evaluation challenges call for an evolution of methodological maturity for the entire community.
We also invite future research to expand on our set of activities and identify challenges unique to specific application domains, such as those addressed in collaborative visualization design studies~\cite{sedlmair2012design}.
For \textbf{practitioners} and \textbf{technology developers}, our paper highlights the shortcomings of remote collaboration tools with respect to supporting data work.
Remote collaboration today means a patchwork of teleconference and productivity tools that altogether limit how participants can contribute to a collaborative activity. 
Improving the interoperability, extensibility, and integration of collaboration and visualization platforms while supporting vibrant external developer communities can foster such innovation and allow researchers to evaluate their approaches at scale. 
Finally, for \textbf{policymakers} and \textbf{funding agencies}, the activities described in \cref{sec:scenarios} underscore the central role of remote and synchronous collaboration around data in domains such as public health, climate science, education, and urban planning, all fields where deliberations must be grounded in data and decisions made through inclusive and collective intelligence.
For at a societal and global level, effective remote collaboration in which people can share and discuss data is part of a project of resiliency, particularly in times where international travel is disrupted by environmental emergencies, public health crises, and geopolitical unrest.

Progress in the face of these challenges goes beyond publications.
First, we must build upon each others' open-source prototypes, and on open data from field observations of deployed systems and existing practices. 
Second, we need more international and interdisciplinary collaborations like those which led to this paper, with project meetings, reading groups, and federated seminars mediated by remote collaboration technology, with visualization researchers speaking to social scientists and technologists across other HCI communities including CSCW and accessibility.

In closing, we believe that our interdisciplinary take on the intersection of visualization and collaborative technology will invite further bridge-building between these communities. 
By taking action together, we can design, evaluate, and deploy collaborative systems that are trustworthy, adaptive, inclusive, and ready for the future of remote data work.


\begin{acks}
This work represents the outcome of NII Shonan Meeting \#213 (\textit{Augmented Multimodal Interaction for Synchronous Presentation, Collaboration, and Education with Remote Audiences})~\cite{shonanreport}, held June 24 --- 27, 2024. 

\bstart{Authorship} The first four authors hosted the Shonan seminar; Brehmer, Cordeil, and Hurter served as overall editors of the paper; Büschel, Saffo\footnote{\textbf{Affiliate Disclaimer}: This paper was prepared for informational purposes with contributions from the Global Technology Applied Research Center of JPMorganChase. It is not investment research or advice, nor a recommendation or offer regarding any security, financial product, or service. JPMorganChase and its affiliates make no representation or warranty, express or implied, as to the accuracy or completeness of the information herein and accept no liability for any consequences arising from its use.}, Jasim, and Prouzeau led breakout groups corresponding with Sections~\ref{sec:challenges:ttt} ---~\ref{sec:challenges:evaluation}, respectively; and the remaining authors (listed in alphabetical order by family name) contributed to breakout topic discussions during and after the seminar.

\bstart{Illustrations}
Sheoli Chaturvedi drew the illustrations in
\cref{fig:teaser}; they are available for hire at \href{https://www.sheoli.art/}{sheoli.art}.

\end{acks}

\bibliographystyle{ACM-Reference-Format}




\end{document}